\documentclass[usenatbib]{mnras}
\usepackage{natbib}
\usepackage{graphicx}
\usepackage[usenames, dvipsnames]{color}
\usepackage{verbatim}
\usepackage{rotating}      
\usepackage{txfonts}
\usepackage{pdflscape}

\newcommand{\ha}{$H\alpha$ }

\title[Clues on Arp 142]{
Clues on Arp 142: The Spiral-Elliptical merger}

\author[M. D. Mora et al. ]
{
\parbox[t]{\textwidth} {Marcelo~D.~Mora,$^1$\thanks{Contact e-mail: mmora@astro.puc.cl}, Sergio~Torres-Flores$^2$, Ver\'onica~Firpo$^{3,2}$, Jose A. Hernandez-Jimenez$^{4,5}$,  Fernanda~Urrutia-Viscarra$^{3,2}$, Claudia Mendes de Oliveira$^{5}$}
\vspace*{6pt}\\
$^1$ Instituto de Astrof\'isica. Pontificia Universidad Cat\'olica de Chile. Vicu\~na Mackenna 4860, 7820436 Macul Santiago Chile. \\
$^2$ Departamento de F\'isica y Astronom\'ia, Universidad de La Serena, Av. Cisternas 1200, La Serena, Chile\\ 
$^3$ Gemini Observatory, Southern Operations Center, La Serena, Chile \\
$^4$ Departamento de Ciencias F\'isicas, Universidad Andres Bello, Fernandez Concha 700, Las Condes, Santiago, Chile \\
$^5$ Departamento de Astronomia, Instituto de Astronomia, Geof\'isica e Ci\^encias Atmosf\'ericas da USP, Rua do Mat\~ao 1226, \\
Cidade Universit\'aria, 05508-090, S\~ao Paulo, Brazil
}
\usepackage{pdfpages}

\begin{document}

\maketitle
\begin{abstract}Nearby merging pairs are unique laboratories in which one can study the gravitational effects on the individual interacting components. In this manuscript, we report the characterization of selected H{\sc ii} regions along the peculiar galaxy NGC\,2936, member of the galaxy pair Arp\,142, an E+S interaction, known as ``The Penguin". Using Gemini South spectroscopy we have derived a high enhancement of the global star formation rate SFR\,=\,35.9 M$_\odot$ yr$^{-1}$ probably stimulated by the interaction. Star-forming regions on this galaxy display oxygen abundances that are consistent with solar metallicities. {{The current data set does not allow us to conclude any clear scenario for NGC 2936.}} Diagnostic diagrams suggest that the central region of NGC\,2936 is ionized by AGN activity and the eastern tidal plume in NGC\,2936 is experiencing a burst of star formation, which may be triggered by the gas compression due to the interaction event with its elliptical companion galaxy: NGC\,2937. {The ionization mechanism of these sources is consistent with shock models of  low-velocities of 200-300\,km\,s$^{-1}$.} The isophotal analysis shows tidal features on NGC\,2937: at inner radii non-concentric (or off-centering) isophotes, and  at large radii, a faint excess of the surface brightness profile with respect to de Vaucouleurs law. By comparing the radial velocity profiles and morphological characteristics of Arp\,142 with a library  of numerical simulations, we conclude that the current stage of the system would be about { $50\pm25$\,Myr} after the first pericenter passage.
\end{abstract}


\begin{keywords}
galaxies: interactions, ISM: H{\sc ii} regions 
\end{keywords}


\section{Introduction}

Galaxy interaction is a process that not only transforms and destroys galaxies, it is also a potential way for internal components such as black holes, star-forming regions, and other features to be assembled.  Merging  is one of the key predictions of the cold dark matter model, which is the dominant scenario for galaxy evolution \citep[e.g.][]{White:1978fk,Blumenthal:1984qf,Cole:2000yq,Conselice:2014rz}, where the merging of dark matter halos lead to the merging of the galaxies associated with the halos \citep{Kauffmann:1993ve}, process that it is expected to keep occurring as dark matter haloes continue to merge along cosmic time. The dense environment plays a role that affects mass growth through gravitational harassment or stripping, removing a significant part of the gas content of a galaxy  \citep[e.g.][]{Moore:1996lq}. These processes may explain the mass distribution observed in the well-defined Hubble sequence of galaxy types in the nearby universe \citep{Tasca:2014fp}.

Merging and interacting galaxies are actually ideal places to study the effects of the gravitational encounters in the structure of galaxies. Different authors have studied the kinematics of merging/interacting galaxies \citep[e.g.][]{Amram:2003gf}, the star formation rate properties of these systems \citep[e.g.][]{Ellison:2008rc}, and the morphology of these perturbed objects 
\citep[e.g.][]{Gallagher:2010xe}. These studies have found peculiar kinematics, enhanced star formation activity, and perturbed morphologies. In the same context, \citet{Xu:2010ix} studied a sample of major mergers of galaxy pairs. These authors found that spiral-spiral pairs (S+S) display enhanced star formation rates with respect to a sample of non-interacting systems. The same authors do not find an enhancement in the SFRs of spiral members of spiral-elliptical pairs (S+E). In the context of S+E pairs, \citet{Weistrop:2012jt} studied the minor merger NGC\,4194, thought to be in an advanced stage of evolution which consists of a gas-rich spiral falling onto an elliptical galaxy \citep{Manthey:2008yf}. These authors found different stellar populations in this system (ranging from 10 Myr up to 5 Gyrs), and they argued that part of this population came from the progenitor galaxies. 
\citet{Jutte:2010zp} observed the molecular gas content of the advance S+E merger NGC\,4441. They found a moderate star formation rate (1-2 M$_\odot$ yr$^{-1}$) with no evidence of cold dense cores of ongoing star formation, thus NGC\,4441 was found to be an example galaxy, in which the starburst has already faded. Authors cited above are one of the few studies that focused on the physical properties of star-forming regions located in spiral galaxies belonging to S+E pairs in an early close encounter stage. At early stages, 
gas clouds collide and the star formation is triggered on the spiral galaxy 
forming young objects like star clusters or complexes  of clusters. \citep[e.g.][]{1992AJ....103..691H,1999AJ....118.1551W,2012AJ....144..156K}.
Then, a detailed study of the physical properties of the newly formed regions can give us important information regarding the interaction process that has taken place during the pair interaction. Finally, these young regions can be used to trace the gas phase oxygen abundance distribution of these interacting spiral galaxies. In this work,  we analyze the spectacular interacting pair Arp\,142, focusing on the link between the interaction, the newly formed regions, the gas,  and the star formation feedback.   

The interacting system Arp\,142 \citep{Arp:1966rp} is located at a distance of $\sim$\,100 Mpc (corrected by the influence of Virgo, Great Attractor and Shapley, \citealt{Mould:2000pb}) in the direction of the Hydra constellation. The system is composed by the spheroidal NGC\,2937 and the spiral galaxy NGC\,2936 that displays a complex morphology. 
The latter has been classified as a ring galaxy by \citet{Freeman:1974kk} and \citet{Madore:2009qa}.   
However, \citet{Romano:2008ye} described it as an arc-like galaxy with no evidence of being a twisted ring, suggesting a nearly edge-on view of the system.  \citet{Xu:2010ix}   
report the detection of several infrared bright star-forming regions in the tails of NGC\,2936, suggesting the presence of dust. The peculiar morphology of NGC\,2936  resembles that of a bird head, where the nucleus corresponds to the eye of the bird \citep{Xu:2010ix}. Other authors have suggested that NGC\,2936 mimics a penguin thus the denomination of penguin galaxy.  Visual inspection of Hubble Space Telescope (HST) WFC3-UVIS2 (Hubble proposal ID:12812, PI: Levay) imaging shows that NGC\,2936 and NGC\,2937 are distinguishable and not fully mixed yet {(see  Figure \ref{Figure1})}, hinting an early stage of the merger (or a going on close encounter). Furthermore, the high spatial resolution of the HST image 
shows clearly that arc-like or ring-like forms described by earlier authors is actually due to the North-West tidal plume stems from NGC\,2936 and bending towards the elliptical member of Arp\,142 (NGC\,2937). The former does not show visual evidence of  shells, {but in this work we explore signs of distortion using an isophotal analysis}, while the spiral component  (NGC\,2936) shows the presence of several objects with colors consistent with those of {young star-forming complexes} located in one of the tidal fields of NGC\,2936, hinting a spiral arm under current disruption. Images also suggest a faint feature seen in projection toward the elliptical on the direction of the semi-major axis. In addition, \textit{Sloan Digital Sky Survey} spectra located in the distorted spiral arm of NGC\,2936 shows strong emission-lines, which suggests the presence of young massive stars ionizing the gas.  {{In this work, we perform a spectroscopic analysis in a sample of star-forming regions located in NGC\,2936, which is complemented with high quality archival data for the whole interacting system, with the aim of understanding  how the local star formation and the physical conditions of the ionized gas were affected -or not- by the interaction. In addition, we discuss the global picture of the merging scenario for this object, which adds important pieces to understand the localized star formation in Arp\,142. Of course, a detailed analysis of localized star-forming regions can be affected by sample effects (e.g. sample size, selection bias, etc), however, very fruitful insights can be obtained from a system that has been not studied in detail in the past.}}
This work is structured as follows: in section \ref{Data} we present the observations, data reduction procedures, and object selection. In section \ref{Analysis} we present the data analysis, in section \ref{Discussion} we discuss our result and finally in section \ref{Conclusions} we draw our conclusions.


\section{Observation and data reduction}
\label{Data}

Observations were acquired in queue mode using the Gemini Multi-Object Spectrograph (GMOS) mounted at Cerro Pach\'on, Gemini South Telescope (program ID: GS-2014B-Q77, PI: Mendes de Oliveira). We have used the $B600$ disperser with a slit width of 1$\arcsec$ pointed into 4 places (see Figure~\ref{Figure1}): near the center of the NGC\,2936 {(Slit\,2)}, across the spiral arms {(Slit\,1 and Slit\,3),} and along several star-forming regions located across the southern tidal tail of NGC\,2936 {(Slit\,4)}. For each pointing, a set of 3 exposures of 900\,sec were acquired including a shift of 5{\AA}~in the central wavelength aimed to avoid the Hamamatsu CCD gaps. Each science observation was then followed by its corresponding arcs and flats aiming at minimizing wavelength differences due to telescope flexures. {The spectral range covered by the observations was from 4200{\AA} to 7400 {\AA}}. The typical seeing was consistent with the requested weather conditions ($\sim$ 1\arcsec) and it was of the order of 1\arcsec. {The spectral resolution of the observations was 4.5 {\AA}, as measured from the full width at half-maximum (FWHM) on the 5557 {\AA} sky line.}

Data reduction was done using Gemini GMOS routines within {\sc pyraf}\footnote{PYRAF is a product of the Space Telescope Science Institute, which is operated by AURA for NASA.}/{\sc iraf}\footnote{Image Reduction and Analysis Facility, distributed by NOAO, operated by AURA, Inc., under agreement with NSF}.  Raw images were bias and flat corrected. Cosmic contamination of the CCDs were removed using the {\tt La Cosmic} routine \citep{van-Dokkum:2001fk} which was applied to individual images. Unfortunately, our images were affected by the hot column problem of the Amplifier 5 in the Central CCD Hamamatsu, therefore we have developed a script to correct it.  The code is available at github\footnote{\href{https://github.com/mmorage/Scripts.-/blob/master/Amp5\_v3\_END.py}{https://github.com/mmorage/Scripts.-/blob/master/Amp5\_v3\_END.py}}. {{Flats and science were quantum efficiency corrected.}} Wavelength calibration was done following the Gemini {\sc pyraf/iraf} package and the typical dispersion was {0.02 {\AA}}. Sky was later subtracted and images were finally combined into a master science frame. {The standard star {{LTT\,1020}} {was observed for flux calibration purposes, with an exposure time of 60\,sec}} and it  was reduced following Gemini standard procedures. 
In addition, HST images (WFC3-UVIS2) used in this work (filters $F475W$, $F606W$, $F675N$, and $F814W$ with exposures times of 2745 sec, 1500 sec, 3600 sec, and 2640 sec respectively) were downloaded from the HST archive.


\begin{table}
\scriptsize
\centering
\caption{Identification, coordinates and measured radial velocities.}
\begin{tabular}{cccc}
\hline
ID & RA(J2000)  & Dec(J2000) & V$_r$ \\
   &            &            &kms$^{-1}$ \\
\hline
S1A  		& 9:37:45.519 & +2:45:32.16 & 7269$\pm$23 \\
S1B  		& 9:37:44.992 & +2:45:33.59 & 7162$\pm$37\\
S1C  		& 9:37:44.696 & +2:45:34.29 & 7155$\pm$16\\
S1D  		& 9:37:44.037 & +2:45:36.08 & 6859$\pm$29\\
S1E  		& 9:37:43.260 & +2:45:38.09 & 7111$\pm$84\\
S2A  		& 9:37:44.051 & +2:45:43.93 & 7208$\pm$42\\
S2B  		& 9:37:44.177 & +2:45:41.25 & 7362$\pm$41\\
S2C 		& 9:37:44.256 & +2:45:39.49 & 7052$\pm$39\\
S2D  		& 9:37:44.458 & +2:45:35.23 & 7017$\pm$66\\
S3A  		& 9:37:44.218 & +2:45:44.06 & 7355$\pm$25\\
S3B        	& 9:37:44.133 & +2:45:43.96 & 7348$\pm$25\\
S3C         & 9:37:43.900 & +2:45:43.74 & 7157$\pm$48\\
S3D         & 9:37:43.805&  +2:45:43.64 & 7113$\pm$49\\
S3E$^{\dag}$& 9:37:43.644 & +2:45:43.00 & 7097$\pm$25\\
S3F  		& 9:37:43.558 & +2:45:43.28 & 7108$\pm$31\\
S4A$^{\dag}$& 9:37:43.260 & +2:45:38.09 & 7101$\pm$28\\
S4B1 		& 9:37:43.643 & +2:45:42.02 & 6752$\pm$1\\
S4B2 		& 9:37:43.643 & +2:45:42.02 & 6760$\pm$2\\
S4C  		& 9:37:43.520 & +2:45:21.13 & 6860$\pm$13\\
S4D  		& 9:37:43.460 & +2:45:10.84 & 6878$\pm$26\\
S4E 		& 9:37:43.451 & +2:45:08.45 & 6960$\pm$14\\
S4F  		& 9:37:43.438 & +2:45:06.74 & 7005$\pm$7 \\
S4G  		& 9:37:43.430 & +2:45:05.23 & 7045$\pm$56\\
\hline 
\label{tableradec}
\end{tabular}
\\{\footnotesize $^{\dag}$Same star-forming region observed in two different slits (S3 and S4).}\\

\end{table}


\subsection{Selection of the individual regions}

\begin{figure}
\includegraphics[width=8.5cm]{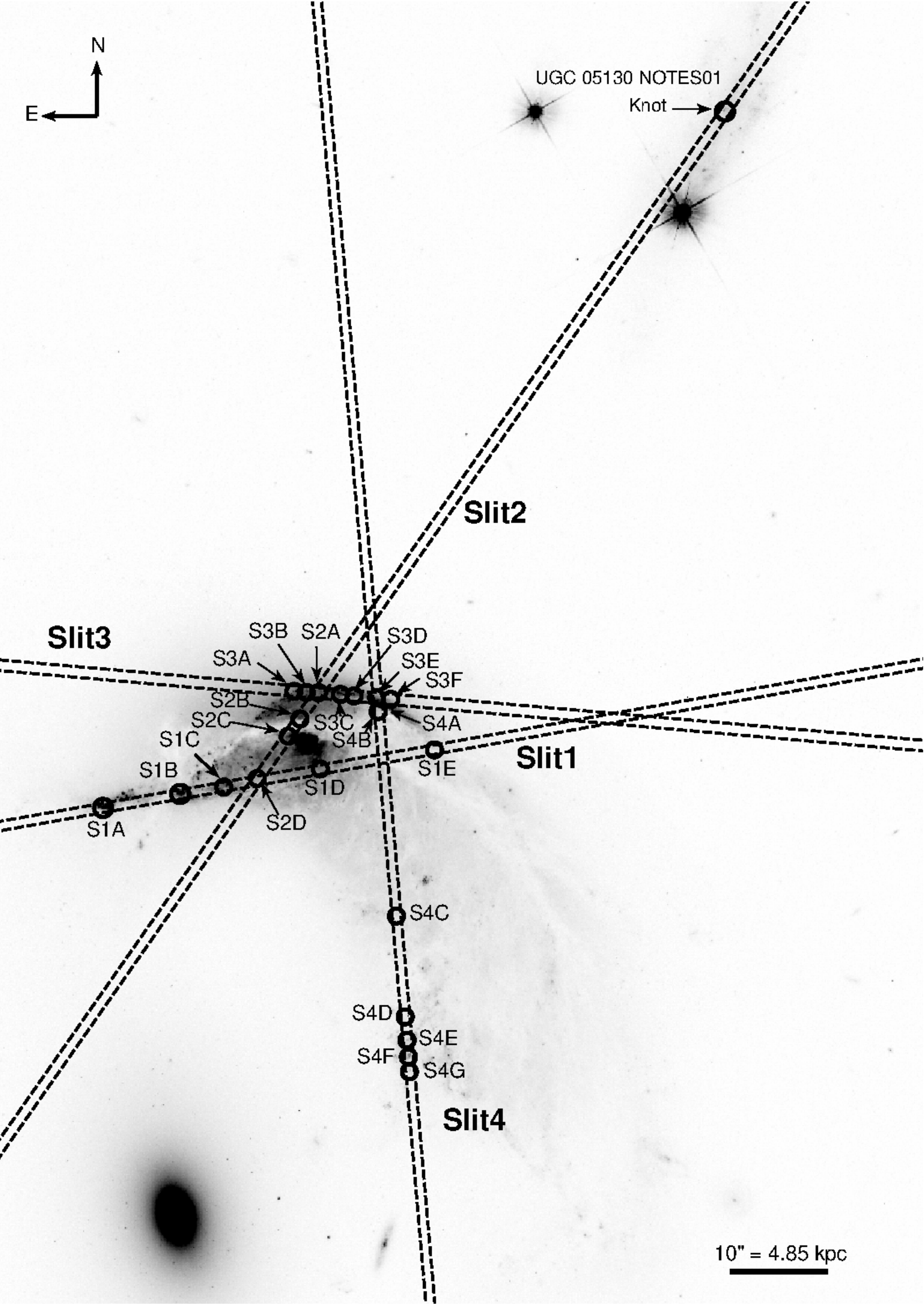}
\caption{{HST $F606W$ image of NGC\,2936.
Dashed lines represent 1" GMOS slit pointings. Circles correspond to the selected H{\sc ii} regions analyzed in this manuscript. Labels were assigned according the target position on their corresponding slits.}}
\label{Figure1}
\end{figure}

{{Since star cluster complexes  are brighter in blue passband than in the redder ones, we used the HST images to compose a color image using the $F475W$ (blue), $F606W$ (green) and, $F814W$ (red) passband, selecting the blue knots that hints to be located in star-forming regions which maximizes the number of object to be observed on each slit. Owing to the seeing limited observations,}} star-forming regions in NGC\,2936 appear as diffuse objects across the different slits. {{Therefore we used the g-filter GMOS-South acquisition image to cross identify individual regions on the 2D spectra and we extracted them by considering the area where the H$\alpha$ profile, measured in the spatial direction on the 2D spectra, intersects the background noise.}}

\begin{figure}
\includegraphics[width=8.5cm]{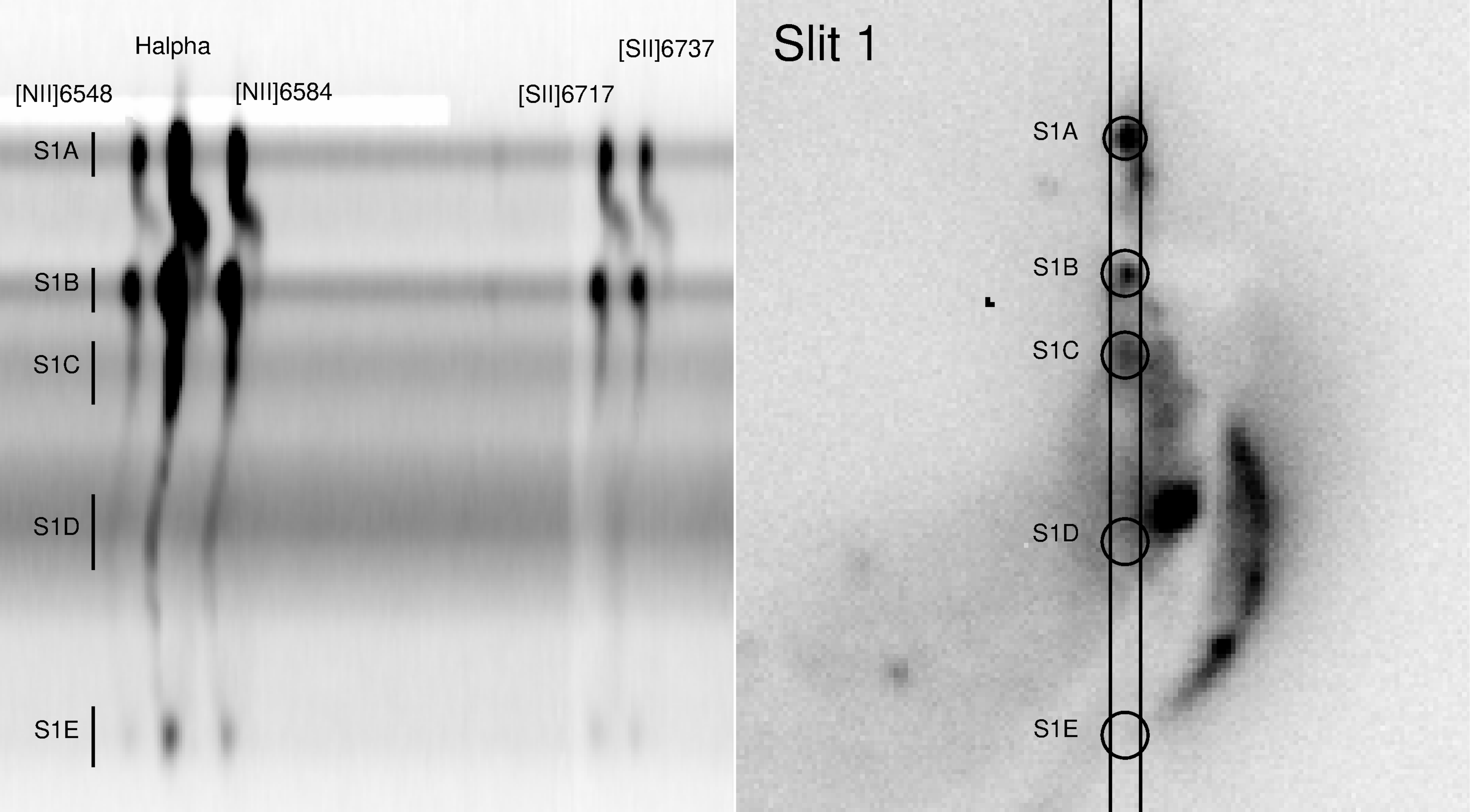}
\includegraphics[width=8.5cm]{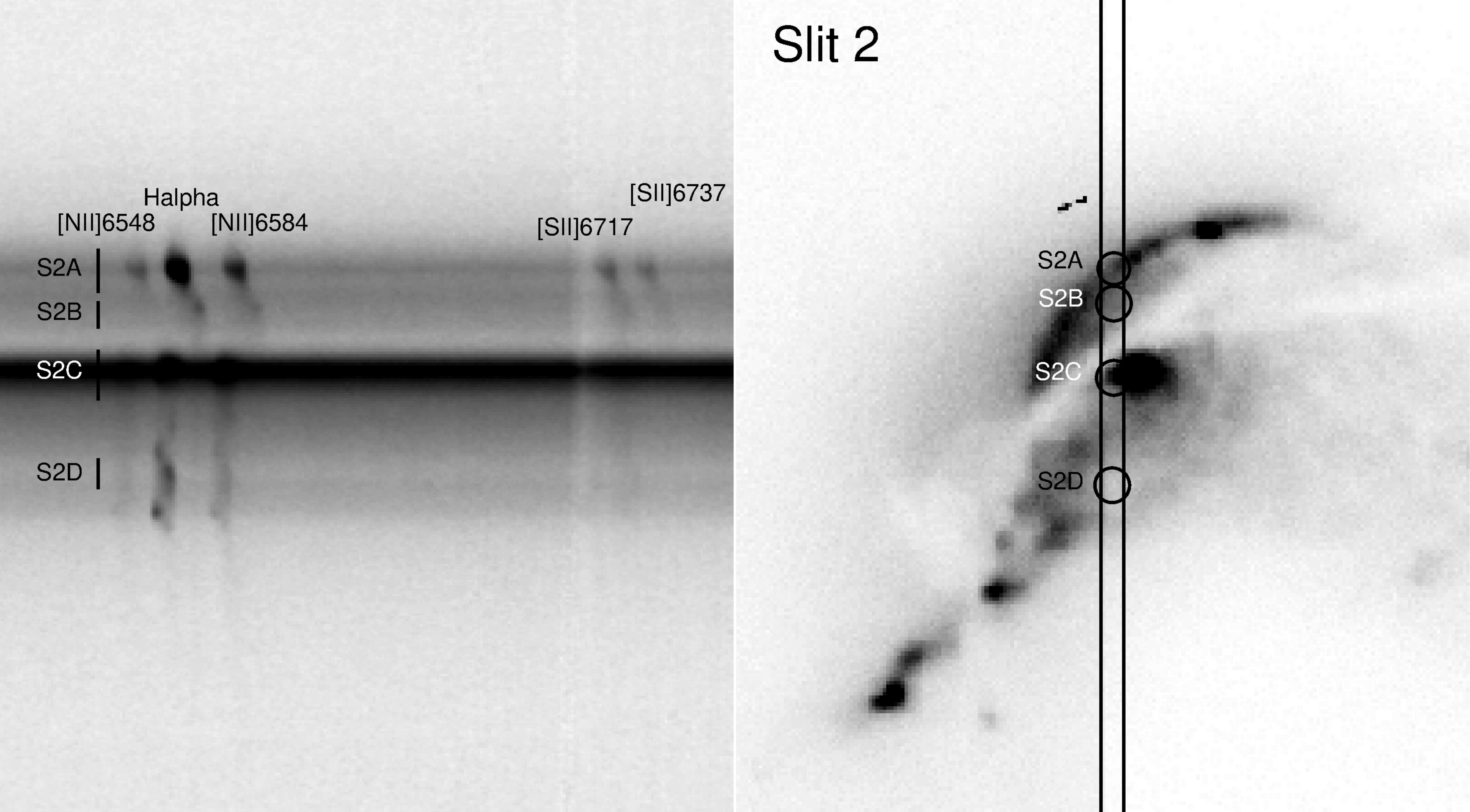}
\includegraphics[width=8.5cm]{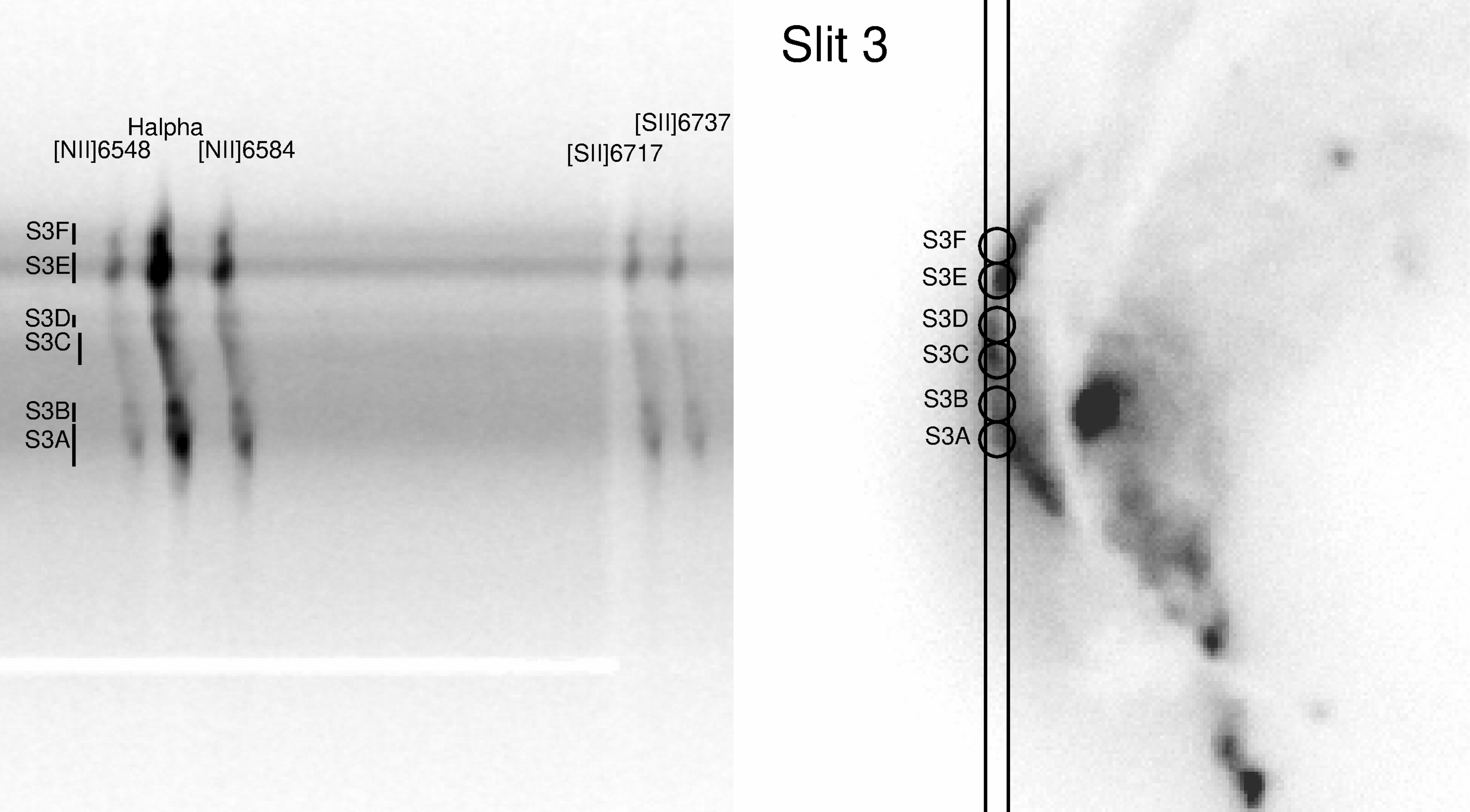}	
\includegraphics[width=8.5cm]{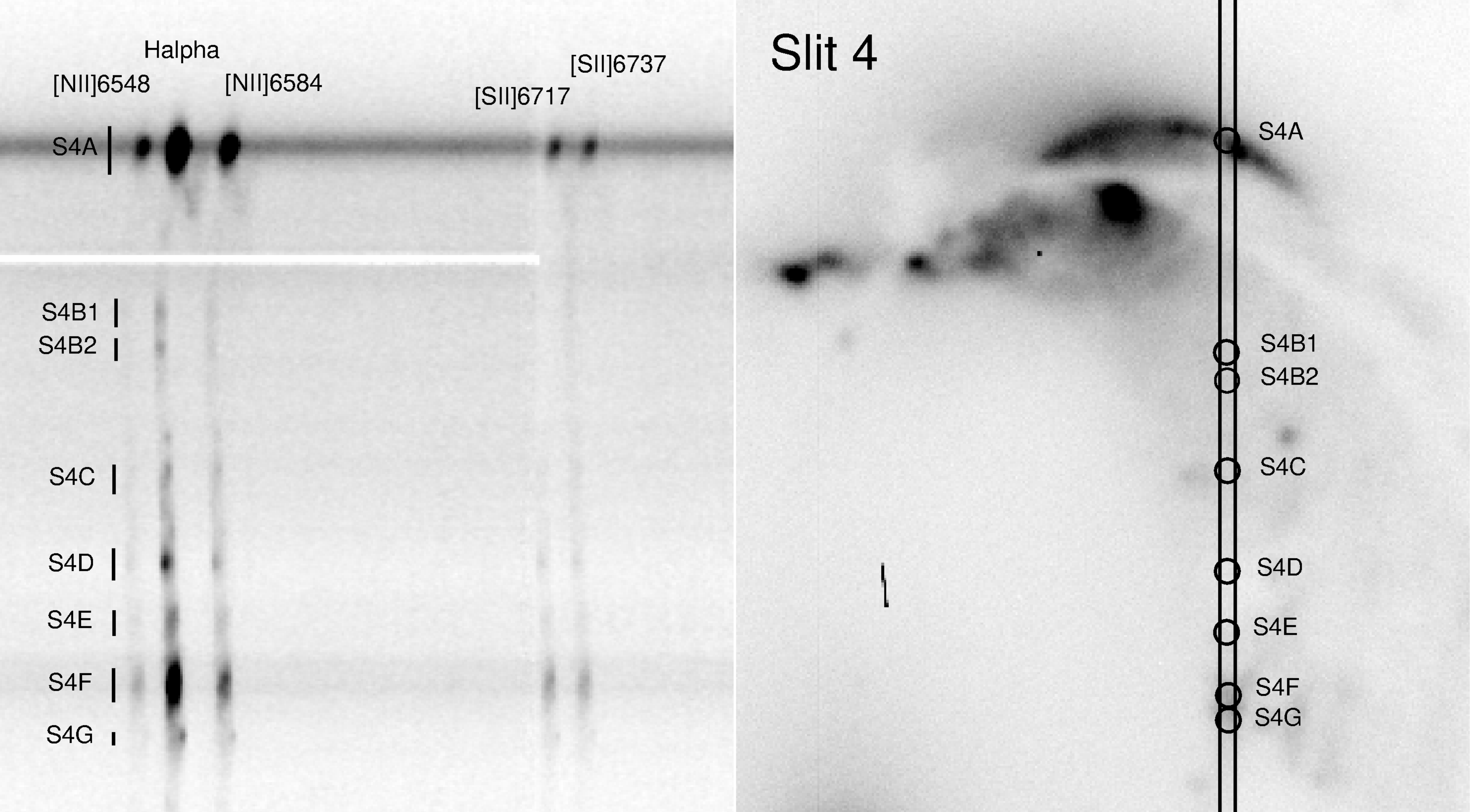}
\caption{{2-D spectra and their corresponding positions in NGC\,2936. On the left panels the} {extracted regions are shown in the zone where H$\alpha$, [N{\sc ii}]$\lambda$6548,84\AA{}, and [S{\sc ii}]$\lambda$6717,37\AA{} are located. Black lines correspond to our selection of the individual regions based on the \ha emission-line. Right panels show $g$-filter GMOS-South acquisition images with the overlapped slits and the selected regions.}}
\label{Figure2}
\end{figure}

These criteria yielded a total of 22 extracted regions.  We have labeled our star-forming region candidates following the order that each region occupies in their respective slit, e.g. S1A is the first region that appearing in the Slit\,1, S1B is the second region, and so on. One of 22 regions was extracted in two different slits (S3E and S4A). In Table \ref{tableradec} we list the identification (ID), RA, Dec, and radial velocities for the selected regions. In  Figure~\ref{Figure1} we present the Wide Field Planetary Camera\,3 image acquired on the $F606W$ passband, downloaded from the Multimission Archive at the Space Telescope Science Institute (MAST). {Circles represent the identified star-forming regions observed in NGC\,2936.} 
In Figure~\ref{Figure2} we present the extended spectra (on the left panels) where H$\alpha$, [N{\sc ii}]$\lambda$6548,84\AA{}, and [S{\sc ii}]$\lambda$6717,37\AA{} are located, while on the right panels we show  the slit positions and the identified knots overlaid on the $g$-band GMOS-South acquisition images.

{Finally, spectra were corrected by Galactic extinction assuming a color excess of E(B-V)\,=\,0.036 \citep{Schlegel:1998qc}  and the \citet{Fitzpatrick:1999ca} extinction law.}


\section{Analysis}
\label{Analysis}

In this section, we analyze the spectroscopic data for NGC\,2936, the spiral galaxy of the Arp\,142 pair. We also derive the radial velocity of the galaxy UGC\,05130\,NOTES01, located  1.3 arcmin north of  the center of NGC\,2631, in projected distance, which is of interest, given that this will give an indication of membership to the system or not. A photometric analysis of the elliptical galaxy  NGC\,2937  is also performed, in order to look for isophotal distortions or faint ripples/shells, indicators of interaction. Finally, a GalMer model simulation of the system is attempted and velocity curves of model and observations are compared.   

\subsection{Emission-line fluxes derived from long-slit spectra}

\begin{figure*}
\includegraphics[width=\textwidth]{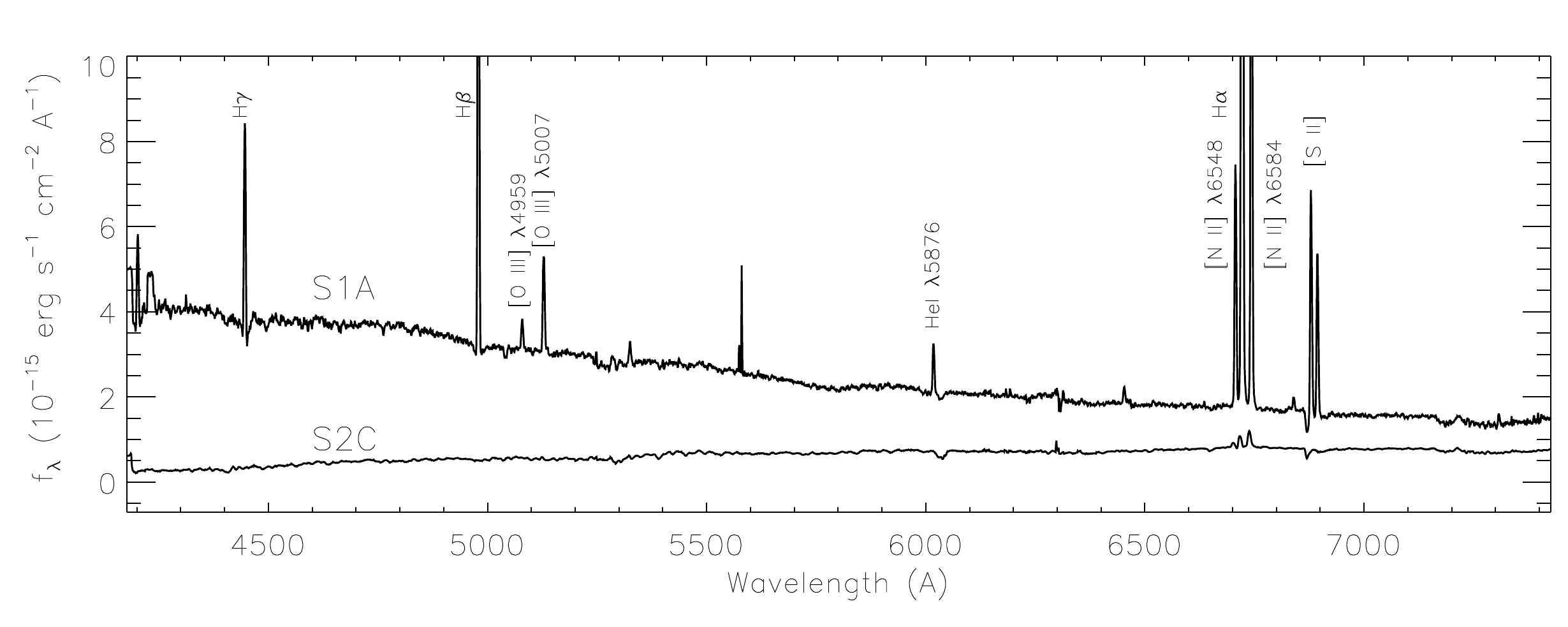}
\includegraphics[width=\textwidth]{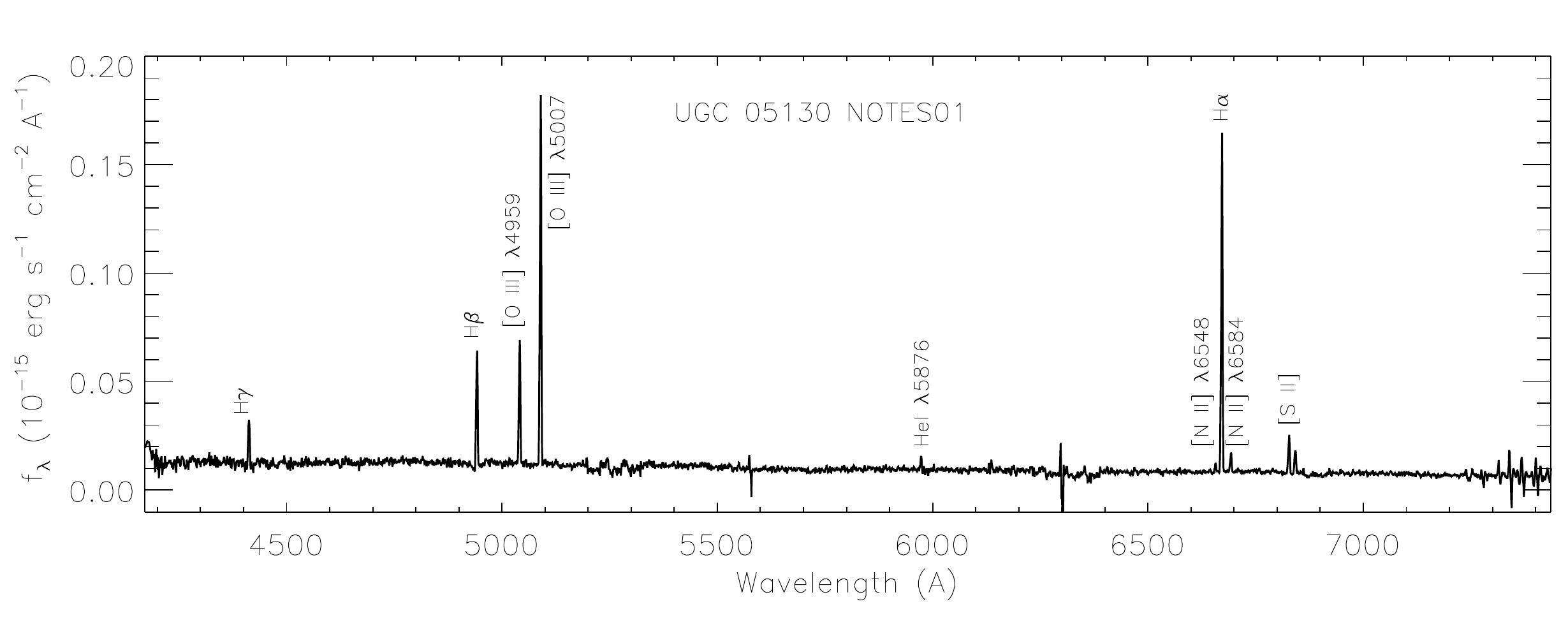}
\caption{Top panel: Typical observed spectra in NGC\,2936. The examples correspond to the region namely S1A and S2C (black continuous lines). Main nebular emission-lines are labeled. Bottom panel: UCG\,05130\,NOTES01 spectrum including the detected emission-lines.} 
\label{Figure3}
\end{figure*}

Long-slits placed in NGC\,2936 were oriented to maximize the number of spectra of the star-forming regions located across NGC\,2936.  
Visual inspection of individual spectra show strong nebular emission-lines, the presence of a weak continuum emission which could be associated with the young stellar population of each star-forming object and the underlying stellar population of NGC\,2936 (except slit S4 where we did not detect continuum).
{{For example, in Figure~\ref{Figure3}, at the top panel, we show the spectrum of one source identified in NGC\,2936 as S1A (black solid line), and in the bottom panel we show the spectrum of the galaxy UCG\,05130\,NOTES01. In both cases is clear the presence of strong emission lines, which dominate the spectra.}}

{Observed spectra where used to measure the emission-line fluxes by using the {\sc iraf} \texttt{splot} task and the uncertainties were estimated using Monte Carlo simulations for 150 runs, given by the \texttt{nerrsample} parameter of the \texttt{splot} task. Then, for each star-forming region, total extinctions were estimated by comparing the observed  H$\alpha$/H$\beta$ ratio with its theoretical value.} This correction is specially important for the case of the star-forming regions which coincide with the strong dust lane that crosses this galaxy. We assumed a electronic temperature of {T$_{e}$\,=\,10$^4$\,K} and an electron density of {N$_{e}$\,=\,100$cm^{-3}$}, which yields a intrinsic ratio of H$\alpha$/H$\beta$\,=\,2.86 \citep{Dominguez:2013xu}. Considering the nature of the spectroscopic sources (mainly emission-line objects), we have adopted a starburst extinction law \citep{Calzetti:2000fh} to correct each individual object by extinction.

{The spectral range of our observations allows us to detect the most intense emission-lines in each star-forming region spectrum, namely: H$\gamma$, H$\beta$, [O{\sc iii}]$\lambda$4959\AA{}, [O{\sc iii}]$\lambda$5007\AA{}, HeI$\lambda$5876\AA{}, [O{\sc i}]$\lambda$6300\AA{}, [N{\sc ii}]$\lambda$6548\AA{}, H$\alpha$, [N{\sc ii}]$\lambda$6584\AA{}, He{\sc i}$\lambda$6678\AA{}, [S{\sc ii}]$\lambda$6716\AA{}, and [S{\sc ii}]$\lambda$6731\AA{}. The emission-line fluxes were measured on each de-reddened spectrum by fitting a single Gaussian profile through the use of the {\texttt{splot} task in {\sc iraf}}. This procedure provided us the flux and the associated uncertainty for each emission-line, which are listed in Table~\ref{tablefluxes_arp142}.}

\begin{table*}
\begin{minipage}[t]{\textwidth}
\centering
\scriptsize
\caption{Emission line fluxes for regions observed in NGC\,2936}
\begin{tabular}{cccccccccccccc}
\hline
ID & H$\gamma$ &  H$\beta$  & [O{\sc iii}]$\lambda$4959 & [O{\sc iii}]$\lambda$5007 & He{\sc i}$\lambda$5876 &  [O{\sc i}]$\lambda$6300 &       [N{\sc ii}]$\lambda$6548 &      H$\alpha$ & [N{\sc ii}]$\lambda$6584  & He{\sc i}$\lambda$6678 &   [S{\sc ii}]$\lambda$6716  &  [S{\sc ii}]$\lambda$6731  \\
& &\multicolumn{10}{c}{10$^{-15}$ erg cm$^{-2}$ s$^{-1}$ } \\
\hline
S1A	&	42.37	 $\pm$	0.2	&	97.98	 $\pm$	0.2	&	5.71	 $\pm$	0.2	&	17.30	 $\pm$	0.2	&	8.38	 $\pm$	0.2	&	3.07	 $\pm$	0.2	&	32.34	 $\pm$	0.2	&	282.90	 $\pm$	0.2	&	94.66	 $\pm$	0.2	&	1.52	 $\pm$	0.2	&	29.01	 $\pm$	0.2	&	21.56	 $\pm$	0.22	\\
S1B	&	240.50	 $\pm$	2.8	&	554.70	 $\pm$	3.1	&	37.92	 $\pm$	4.6	&	89.01	 $\pm$	3.5	&	56.27	 $\pm$	3.3	&	51.96	 $\pm$	4.7	&	181.80	 $\pm$	2.8	&	1608.00	 $\pm$	2.6	&	534.40	 $\pm$	2.4	&	13.08	 $\pm$	5.7	&	123	 $\pm$	2.6	&	114.1	 $\pm$	2.5	\\
S1C	&	22.99	 $\pm$	0.4	&	47.87	 $\pm$	0.4	&	3.6	 $\pm$	0.4	&	10.85	 $\pm$	0.4	&	5.05	 $\pm$	0.5	&	5.43	 $\pm$	0.4	&	13.40	 $\pm$	0.3	&	136.60	 $\pm$	0.3	&	42.70	 $\pm$	0.4	&	---			&	10.13	 $\pm$	0.3	&	8.44	 $\pm$	0.3	\\
S1D	&	---			&	---			&	---			&	---			&	---			&	---			&	1.06	 $\pm$	0.2	&	5.42	 $\pm$	0.2	&	3.90	 $\pm$	0.2	&	---			&	---			&	---			\\
S1E	&	---	 		&	15.32	 $\pm$	0.3	&	---			&	7.42	 $\pm$	0.3	&	---			&	8.34	 $\pm$	0.4	&	5.97	 $\pm$	0.3	&	43.74	 $\pm$	0.3	&	20.87	 $\pm$	0.3	&	1.21	 $\pm$	0.2	&	4.89	 $\pm$	0.3	&	3.41	 $\pm$	0.2	 \\
S2A	&	11.38	 $\pm$	0.3	&	27.84	 $\pm$	0.2	&	2.59	 $\pm$	0.3	&	5.53	 $\pm$	0.2	&	1.69	 $\pm$	0.3	&	---			&	6.86	 $\pm$	0.3	&	78.67	 $\pm$	0.2	&	22.97	 $\pm$	0.3	&	---		&	4.97	 $\pm$	0.2	&	3.99	 $\pm$	0.3  \\
S2B	&	3.51	 $\pm$	0.2	&	9.50	 $\pm$	0.1	&	0.90	 $\pm$	0.2	&	2.15	 $\pm$	0.2	&	---			&	---			&	2.31	 $\pm$	0.2	&	26.58	 $\pm$	0.2	&	8.15	 $\pm$	0.2	&	---		&	1.90	 $\pm$	0.1	&	1.50	 $\pm$	0.2	 \\
S2C	&	---			&	---			&	---			&	---			&	---			&	---			&	4.65	 $\pm$	0.6	&	13.00	 $\pm$	0.5	&	17.48	 $\pm$	0.4	&	---		&	---			&	---			\\
S2D	&	---			&	18.79	$\pm$	1.0	&	---			&	---			&	---			&	---			&	7.18	 $\pm$	0.8	&	53.08	 $\pm$	0.8	&	19.52	 $\pm$	0.9	&	---		&	6.69	 $\pm$	2.2	&	---	\\
S3A	&	13.33	 $\pm$	0.3	&	33.63	 $\pm$	0.2	&	3.81	 $\pm$	0.3	&	8.21	 $\pm$	0.2	&	---			&	2.51	 $\pm$	0.4	&	9.38	 $\pm$	0.3	&	97.18	 $\pm$	0.4	&	32.05	 $\pm$	0.3	&	---			&	12.76	 $\pm$	0.3	&	8.25	 $\pm$	0.3	 \\
S3B	&	3.78	 $\pm$	0.1	&	8.62	 $\pm$	0.1	&	1.01	 $\pm$	0.1	&	2.17	 $\pm$	0.1	&	---			&	0.67	 $\pm$	0.1	&	2.29	 $\pm$	0.1	&	24.49	 $\pm$	0.1	&	8.00	 $\pm$	0.1	&	---			&	3.43	 $\pm$	0.1	&	2.18	 $\pm$	0.1	 \\
*S3C	&	2.67	 $\pm$	0.1	&	5.77	 $\pm$	0.1	&	0.35	 $\pm$	0.1	&	1.18	 $\pm$	0.1	&	---			&	0.64	 $\pm$	0.2	&	1.65	 $\pm$	0.1	&	15.52	 $\pm$	0.1	&	5.17	 $\pm$	0.1	&	0.25	 $\pm$	0.3	&	1.47	 $\pm$	0.1	&	0.92	 $\pm$	0.1\\
S3D	&	1.14	 $\pm$	0.04	&	2.55	 $\pm$	0.035	&	0.14	 $\pm$	0.04	&	0.45	 $\pm$	0.04	&	---			&	0.19	 $\pm$	0.1	&	0.68	 $\pm$	0.04	&	6.84	 $\pm$	0.045	&	2.23	 $\pm$	0.1	&	---			&	0.54	 $\pm$	0.04	&	0.36	 $\pm$	0.04	 \\
S3E$^{\dag}$	&	38.21	 $\pm$	0.4	&	110.20	 $\pm$	0.4	&	4.34	 $\pm$	0.5	&	14.81	 $\pm$	0.5	&	8.30	 $\pm$	0.5	&	2.67	 $\pm$	0.5	&	26.45	 $\pm$	0.4	&	315.00	 $\pm$	0.4	&	82.95	 $\pm$	0.5	&	---			&	13.67	 $\pm$	0.4	&	13.25	 $\pm$	0.5	\\
S3F	&	23.94	 $\pm$	0.1	&	67.65	 $\pm$	0.2	&	1.87	 $\pm$	0.3	&	8.69	 $\pm$	0.2	&	6.09	 $\pm$	0.2	&	1.63	 $\pm$	0.2	&	15.39	 $\pm$	0.2	&	191.00	 $\pm$	0.2	&	48.65	 $\pm$	0.2	&	---			&	8.50	 $\pm$	0.2	&	8.65	 $\pm$	0.1	\\
S4A$^{\dag}$	&	35.00	 $\pm$	0.2	&	83.50	 $\pm$	0.2	&	3.25	 $\pm$	0.2	&	11.18	 $\pm$	0.3	&	5.49	 $\pm$	0.3	&		---		&	19.46	 $\pm$	0.2	&	239.10	 $\pm$	0.3	&	62.53	 $\pm$	0.3	&	---			&	8.12	 $\pm$	0.3	&	8.64	 $\pm$	0.3	\\
S4B1	&		---		&		---		&		---		&		---		&		---		&		---		&	0.46	 $\pm$	0.05	&	1.68	 $\pm$	0.03	&	0.68	 $\pm$	0.04	&	---			&		---		&	---			\\
S4B2	&		---		&		---		&		---		&		---		&		---		&		---		&	0.18	 $\pm$	0.03	&	0.86	 $\pm$	0.02	&	0.40	 $\pm$	0.03	&	---			&		---		&	---			\\
S4C	&		---		&	1.08	 $\pm$	0.2	&		---		&		---		&		---		&		---		&	1.18	 $\pm$	0.2	&	3.18	 $\pm$	0.2	&	1.74	 $\pm$	0.2	&	---			&		---		&	---			 \\
S4D	&		---		&	13.07	 $\pm$	0.2	&	2.23	 $\pm$	0.3	&	4.31	 $\pm$	0.2	&	1.74	 $\pm$	0.2	&	1.71	 $\pm$	0.3	&	5.38	 $\pm$	0.2	&	36.38	 $\pm$	0.2	&	13.31	 $\pm$	0.2	&	---			&	4.92	 $\pm$	0.3	&	4.35	 $\pm$	0.2	 \\
S4E	&		---		&	2.36	 $\pm$	0.06	&	0.55	 $\pm$	0.09	&	1.22	 $\pm$	0.07	&		---		&		---		&	0.89	 $\pm$	0.07	&	6.75	 $\pm$	0.05	&	2.48	 $\pm$	0.05	&	---			&	0.92	 $\pm$	0.06	&	0.76	 $\pm$	0.06	 \\
S4F	&	5.25	 $\pm$	0.06	&	12.28	 $\pm$	0.06	&	2.98	 $\pm$	0.07	&	9.67	 $\pm$	0.07	&	1.45	 $\pm$	0.08	&	1.17	 $\pm$	0.09	&	3.78	 $\pm$	0.06	&	35.07	 $\pm$	0.06	&	10.56	 $\pm$	0.06	&	---			&	3.99	 $\pm$	0.06	&	3.71	 $\pm$	0.07	 \\
S4G	&		---		&	1.48	 $\pm$	0.06	&	0.42	 $\pm$	0.09	&	0.99	 $\pm$	0.04	&		---		&	0.32	 $\pm$	0.06	&	0.64	 $\pm$	0.08	&	4.26	 $\pm$	0.05	&	1.53	 $\pm$	0.06	&	---			&	0.72	 $\pm$	0.05	&	0.67	 $\pm$	0.05	\\
\hline 
\multicolumn{10}{@{}l@{}}{\footnotesize $^{\dag}$Same star-forming region observed in two different slits (S3 and S4).}\\
\label{tablefluxes_arp142}
\end{tabular}
\end{minipage}
\end{table*}


\subsection{Physical quantities for NGC\,2936}

\subsubsection{Star formation activity }

\label{star_formation_activity}
It is well known that the H$\alpha$ luminosity is a good tracer of the current star formation \citep{Kennicutt:2012pt}, and its emission can be associated with the presence of a large number of young and massive stars whose ultraviolet flux ionize the surrounding gas, where the recombination process is a strong evidence of recent or ongoing massive star formation. Hence, the star-forming regions located in the interacting system NGC\,2936 provide us with a laboratory to study the star formation that may have been triggered by gravitational effects or previously triggered by the internal stimulation of the spiral arms inherent to a galaxy of this type. 
{{In this context, a standard procedure would be the estimation of the current star formation rate by using H$\alpha$ narrow band imaging. However, the redshift of Arp\,142 invalidate this method, given that the H$\alpha$ line is off the F657N \textit{HST} filter.}} Therefore, H$\alpha$ luminosities were calculated using the spectroscopic information by measuring the different H$\alpha$ fluxes and taking into account the distance to Arp\,142.
{{We found  H$\alpha$ luminosities to be in the range from log(L$_{H\alpha}$) $\sim$ 39.08 up to 42.35) erg s$^{-1}$, which is consistent with what have been observed in giant star-forming regions (E.g. \citet{Firpo:2005nf,Firpo:2010fv,Firpo:2011vf}, log(L$_{H\alpha}$) $\sim$ 39.4 erg s$^{-1}$ to 40.9 erg s$^{-1}$) and HII galaxies (e.g. \citet{Hagele:2008gf}, log(L$_{H\alpha}$) $\sim$ 38.9 erg s$^{-1}$ to 41.6 erg s$^{-1}$)}.}  {{We then converted these luminosities into star formation rates (SFRs)  using the classical equation given in \citet{Kennicutt:1998ly}, which assumes a continuous star formation}} {{process over the timescale associated with the star-forming tracer. This calibrator considers a Salpeter Initial Mass Function, with masses ranging from 0.1-100 Msun.}} {{We derive SFR that are in agreement with values found by \citet{Ferreiro:2008cs} in the star-forming regions studied in interacting galaxies (0.07 to 10 M$_\odot$ yr$^{-1}$).}}
Finally, results are  listed in Table~\ref{tableabundances_arp142}, where we list the H$\alpha$ luminosities and SFRs measured for each region belonging to NGC\,2936 as well as their values with no reddening corrections (H$\alpha_{uncorr}$ and SFR$_{uncorr}$), as well as their measured excess colors [E($B-V$)]. We also include the star formation rate density for each star-forming region ($\Sigma_{SFR}$), which was derived as the SFR divided by the area within which the spectra were extracted.

\begin{figure*}
\includegraphics[scale=0.45]{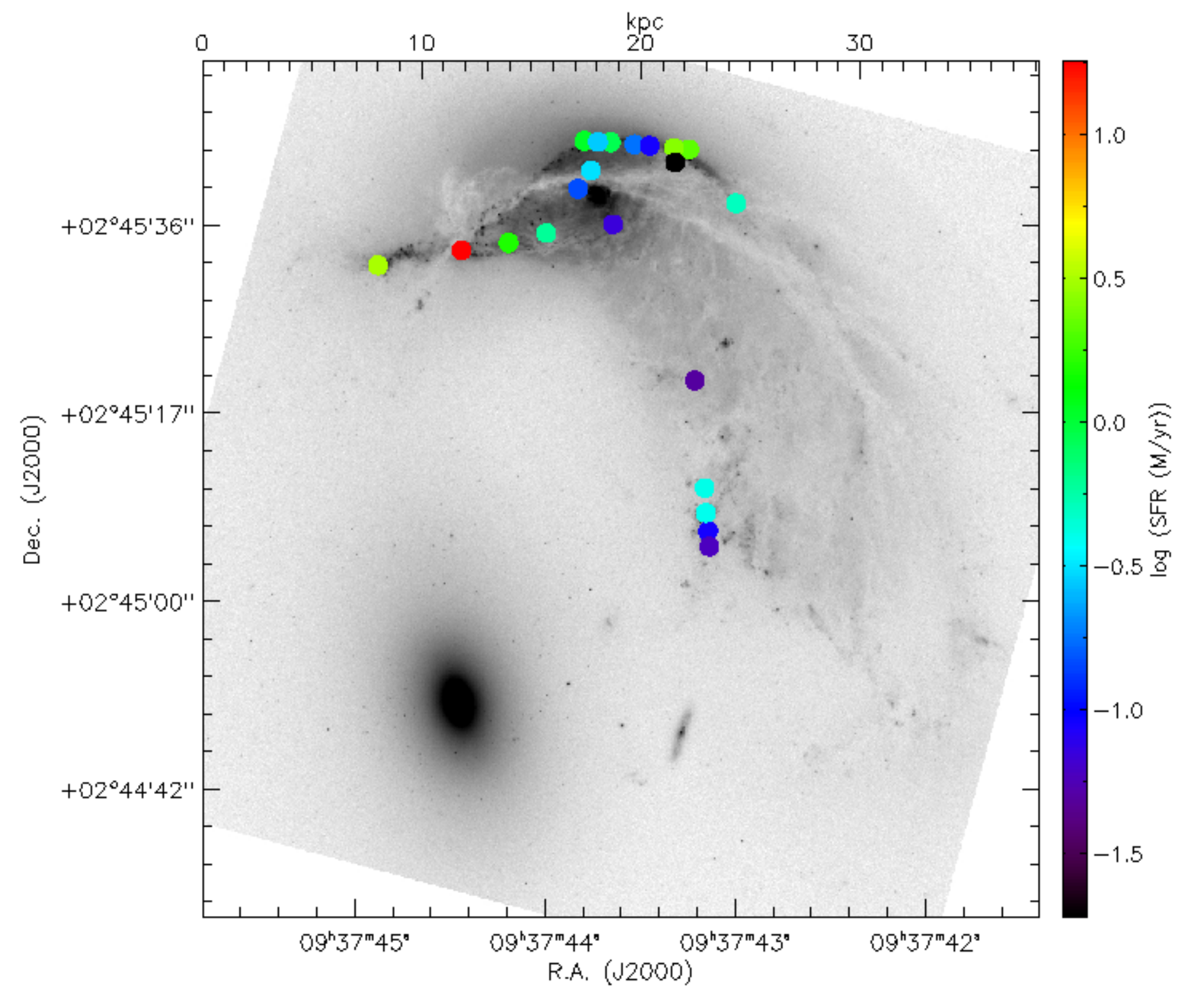}
\includegraphics[scale=0.45]{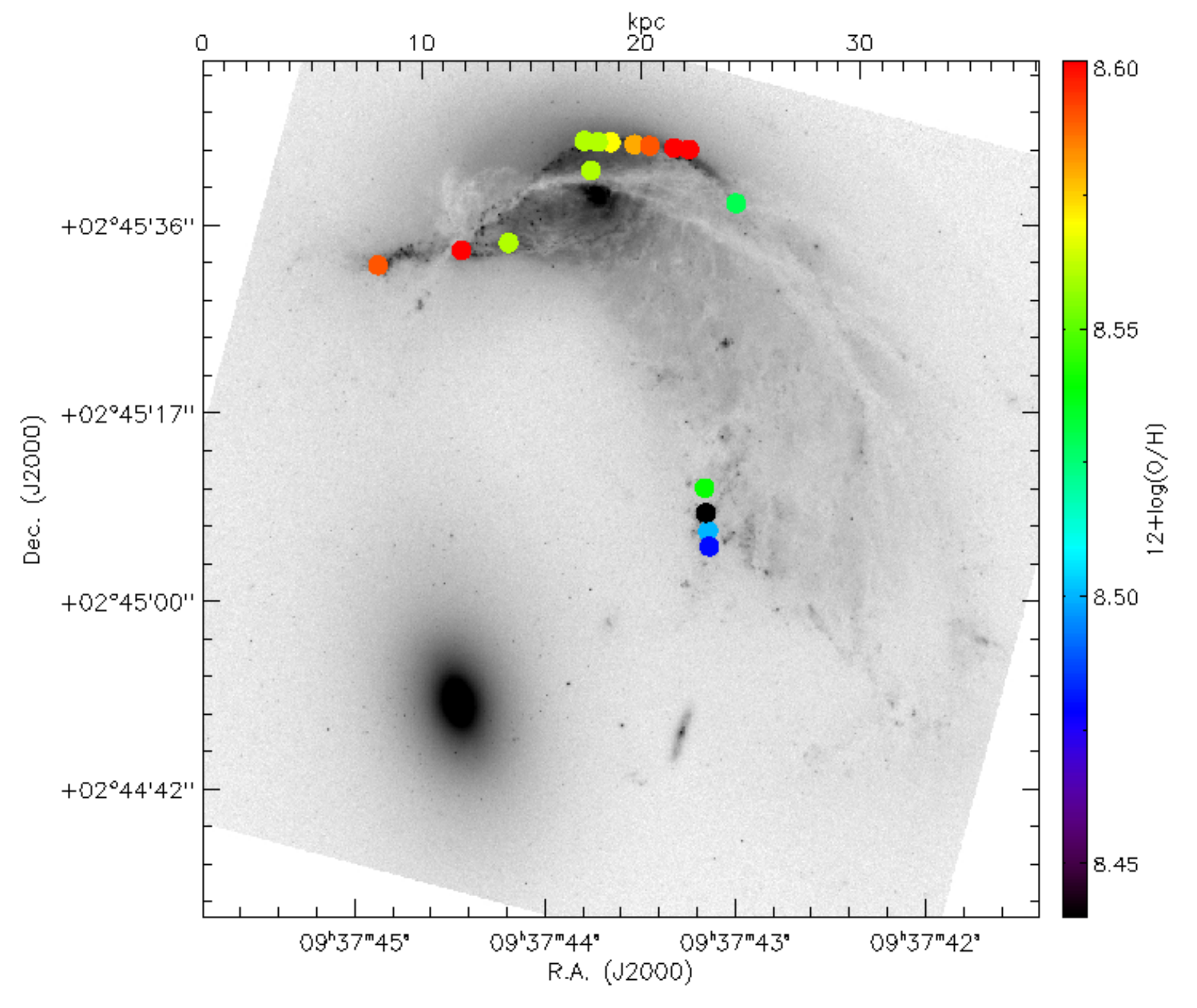}
\includegraphics[scale=0.45]{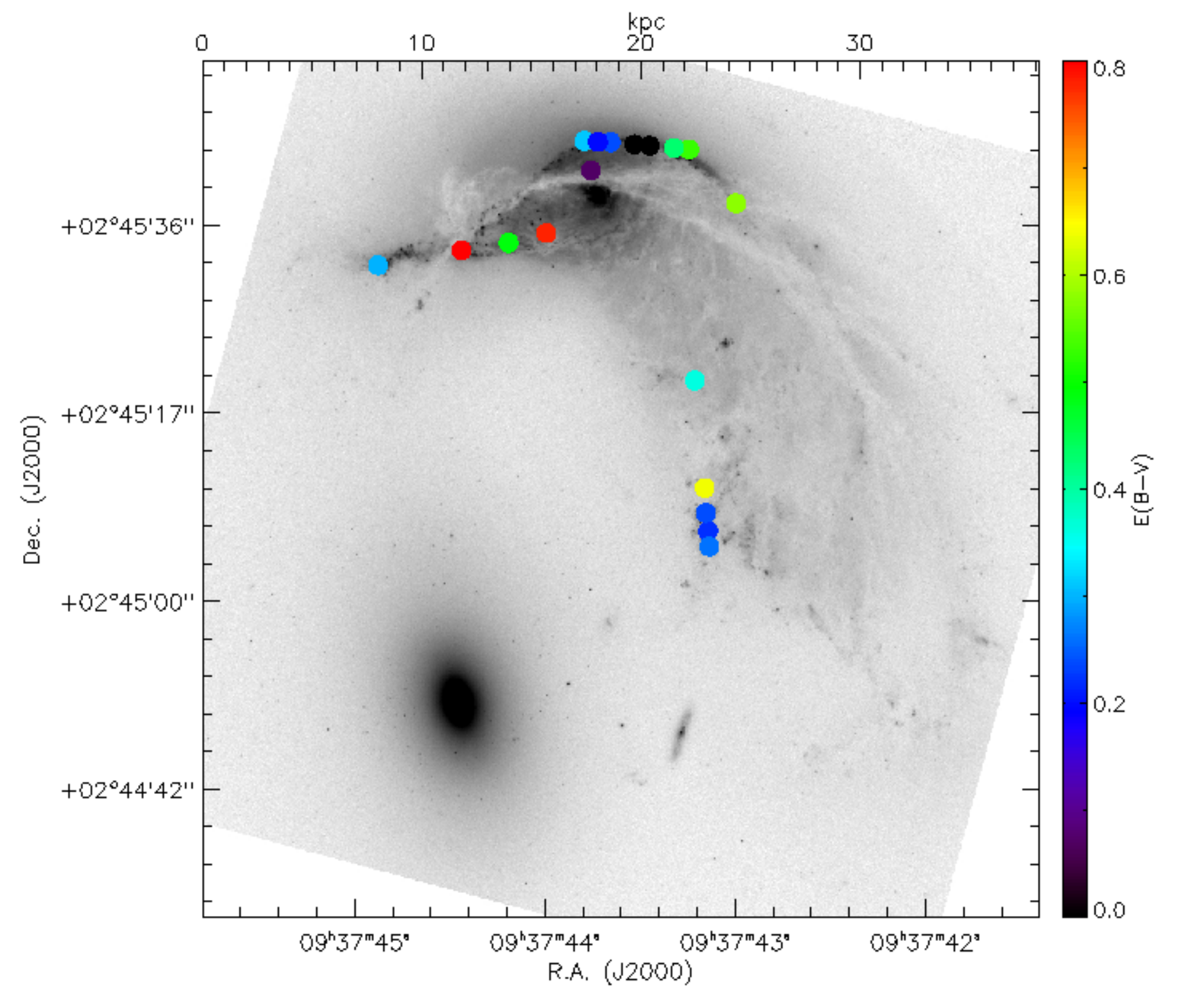}
\includegraphics[scale=0.45]{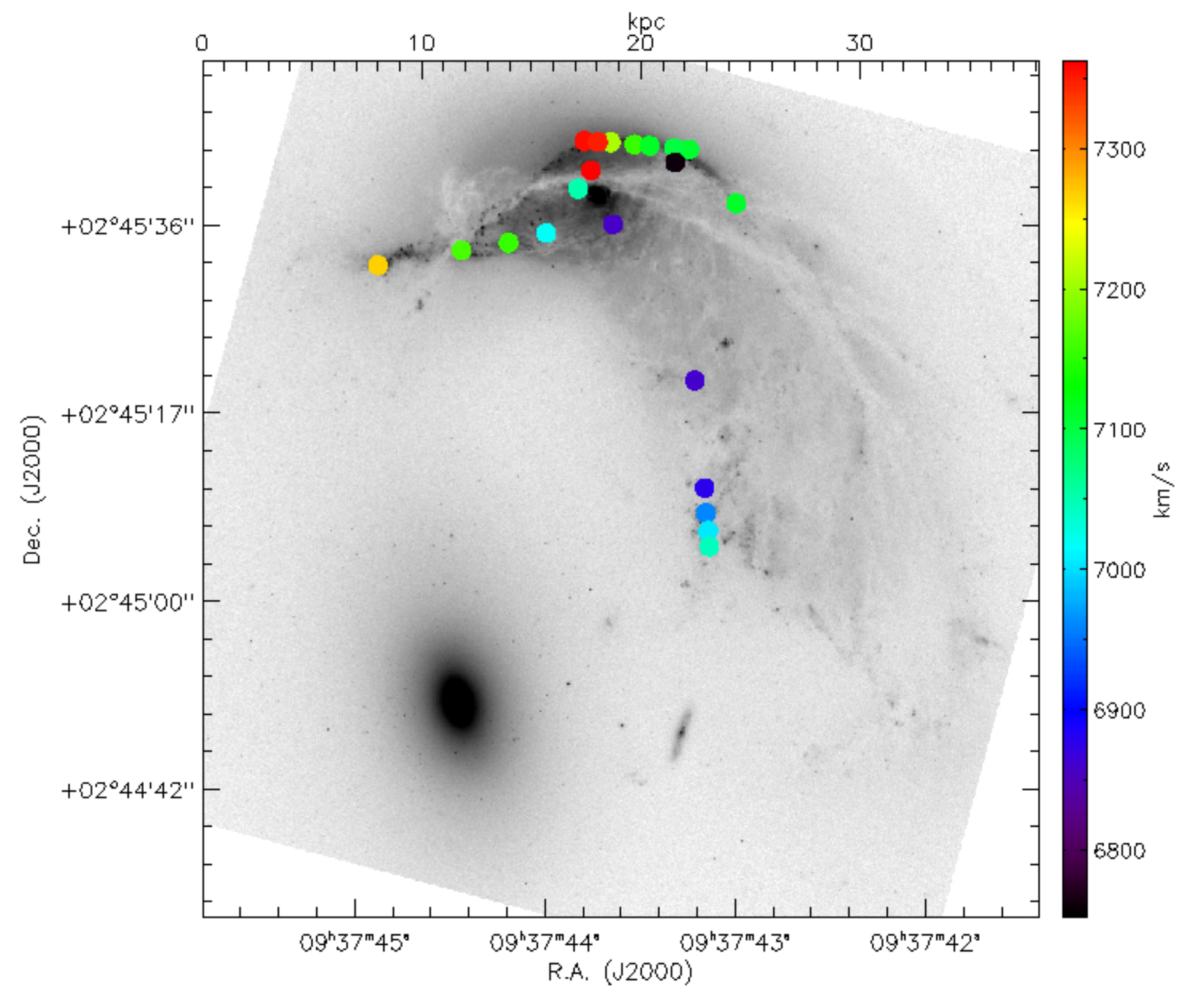}
\caption{Color-coded spatial distribution of the H{\sc ii} regions on top of the NGC\,2936 HST image. On the top left panel, we present the SFRs. On the Top right panel, we present the oxygen abundances. On the bottom left, the extinction, and on the bottom right the H{\sc ii} velocity measurements. Circle sizes do not represent physical quantities and were expanded for best representation in the figure.}
\label{Figure4}
\end{figure*}

In the top left panel of Figure~\ref{Figure4} we present the spatial distribution of the star formation rates in the regions of NGC\,2936. Although this plot does not have the spatial resolution to show localized variations of the quantities, it can give us an idea of trends and variations seen in our data. From this plot, we can see that NGC\,2936 is actively forming stars in its eastern side. Region S1B displays the highest SFR, which corresponds to 17.9 M$_\odot$ yr$^{-1}$. This source lies over a dust lane, which produces a high extinction at this location (see section \ref{extinctions+electron}). The second most intense star formation burst corresponds to S1A, which is located at the tip of the eastern tidal structure in NGC\,2936. This object has a SFR of 3.2 M$_\odot$  yr$^{-1}$. Both values are extremely high when compared with typical SFRs derived for spiral galaxies \citep[see][]{2007ApJ...658.1006M}. Indeed, the star formation rate densities of these sources correspond to log($\Sigma_{SFR,cor}$)= 1.30 M$_\odot$ yr$^{-1}$ kpc$^{-2}$ and log($\Sigma_{SFR,cor}$)=0.50 M$_\odot$ yr$^{-1}$ kpc$^{-2}$, which reflects that both objects correspond to a localized starburst \citep{Efremov:2004bs}. In the case of the first source, its $\Sigma_{SFR,cor}$ is consistent with the value shown by star-forming regions located in LIRG galaxies \citep[e.g.][]{Piqueras-Lopez:2016qr}. However, these estimates could be overestimated. In this analysis we have used the \citet{Calzetti:2000fh} extinction law, which was defined for starburst objects. In order to perform some comparisons, we have dust-corrected the spectrum of region S1B by using the \citet{Fitzpatrick:1999ca} extinction law. In this case, we have derived a SFR=8.7 M$_\odot$  yr$^{-1}$ for this source, which is much lower than the value derived from the spectrum corrected by using the \citet{Calzetti:2000fh} extinction law. This comparison shows how sensitive are our results depending on the use of a specific extinction law.

In any scenario, these results may suggest that the eastern tidal structure/arm in NGC\,2936 is experiencing a burst of star formation. Despite that the current data set does not allow us to conclude that the burst was triggered by gas compression during the interaction in Arp\,142, there are evidence that suggests that interacting galaxies display strong burst of star-formation. For instance, \citet{Ferreiro:2008cs} found bright H$\alpha$ knots in a sample of interacting galaxies. \citet{Ellison:2008rc} found that galaxy pairs (at separation lower than 30-40 h$^{-1}$ 70 kpc) display an enhancement in their SFRs. Therefore, based in our results, we can speculate that the strong star formation bursts that we are witnessing in NGC\,2936 are the result of the gravitational encounter between the main members of Arp\,142. This encounter have produced an enhancement in the SFR of this galaxy pair, which is mainly focused in S1B. Actually, when we take into account global values, we obtain a total SFR of 35.9 M$_\odot$ yr$^{-1}$ for NGC\,2916 (which could be $\sim$17.5 M$_\odot$ yr$^{-1}$ if we scale the value for the \citet{Fitzpatrick:1999ca} extinction law). This value decreases 7.35 M$_\odot$ yr$^{-1}$, if we consider the fluxes that have not been corrected by internal extinction. Both values reflect a prominent SFR in this galaxy pair.
{In addition, regions S3E, S3F and S4A display high values for the star formation rate, where all these sources are located in a spiral structure.} All the remaining sources have considerably low SFRs with respect to the former objects.


\subsubsection{Gas phase abundances}
\label{Gas-abundances}
Recent studies on the distribution of chemical abundances in interacting galaxies have provided important clues to understand the metal/gas mixing process in these systems \citep{Kewley:2006pi,Rupke:2010bc,Barrera-Ballesteros:2015im}. Most of these works rely on  the oxygen abundance to study the chemical content of galaxies. However, due to the difficulty to observe weak auroral lines sensitive to temperature (such as [O{\sc iii}]$\lambda$4363\AA{} which allow determining  oxygen abundances directly), motivated us to use  empirical calibrations through the O$_3$N$_2$  method \citep{Pettini:2004lp}. This is one of the most used methods involving strong emission-lines  such as the H$\beta$, [O{\sc iii}]$\lambda$5007\AA{}, H$\alpha$, and [N{\sc ii}]$\lambda$6584\AA{} lines. Given the emission-lines available in our data, we have estimated oxygen abundances by using the updated O$_3$N$_2$ calibration proposed by \citet{Marino:2013nz}. 
The typical dispersion on this calibration corresponds to 0.18 dex. Therefore, uncertainties in the oxygen abundances are the quadratic sum of the uncertainties in the flux and the dispersion of the calibrator. {Finally, derived oxygen abundances, listed in Table~\ref{tableabundances_arp142}, show values between 8.44-8.60$\pm$0.18, and (within the uncertainties) are consistent with the solar abundance   \citep[12+$\log$(O/H)$_{\odot}$\,=\,8.69,][]{Allende-Prieto:2001zh}. These estimated values are in agreement, within the uncertainties, with that found by \citet{Olave-Rojas:2015wj}(8.11$\pm$0.18 to 8.53$\pm$0.17) and \citet{Rosa:2014al} (8.54$\pm$0.01 to 8.83$\pm$0.02).}

{{In the right top panel of Figure~\ref{Figure4}  we show the spatial distribution of the oxygen abundances in the regions of  NGC\,2936. We found that four star-forming regions located in the southern tidal structure of NGC\,2636 have a slightly lower oxygen abundances than the remaining objects {(12+log(O/H)\,$\sim$\,8.45-8.52). Considering that the tail structure could be formed from material belonging to the outskirts of NGC\,2936, it could be expected to observe a lower oxygen abundance at this location compared with the central part of the galaxy. However  the very localized sample presented here  and the  corresponding uncertainties do not allow us to fully rule out its absence in the main body of the galaxy  as it  have been observed in a similar way  in other strongly interacting systems  such as  NGC\,4676 \citep[][]{Chien:2007qz} and  NGC\,92 \citep{Torres-Flores:2014bs} and in ongoing interactions \citep[e.g.][]{Rupke:2010bc}.}}}


\subsubsection{Extinctions and electron densities}
\label{extinctions+electron}
By using the H$\alpha$/H$\beta$ Balmer ratio we obtained the internal extinction for all the sources. The spatial distribution of these values is shown in the bottom left panel of Figure~\ref{Figure4}. Three knots (S1B, S2C and, S4D) show high extinction, with values ranging from E(B-V)\,=\,0.67  up to E(B-V)\,=\,0.74. Two of them are located over the inner dust lane of the Eastern arm. 
The third one is located close to NGC\,2936 nucleus, in a region where dust is seen spread out throughout the optical image. Although the map shown in the bottom left panel of Figure~\ref{Figure4} displays a discrete distribution of points, it reveals the critical influence of the dust lane on the physical properties of the star-forming regions located in this system.

Electron densities (N$_{e}$) represent the low-excitation zone of the ionized gas, and these were derived from the parameter R$_{S2}$, which is defined as the ratio between the [S{\sc ii}]$\lambda$6717/6731\AA\ emission-lines. We used the \texttt{temden} routine of the {\sc nebular} package in {\sc STSDAS/iraf}, assuming an electron temperature of T$_{e}$\,=\,10$^4$\,K.  The calculated N$_e$ values are given in Table \ref{tableabundances_arp142}, from them, {we can see that in the eastern region of NGC\,2936 the estimated values of electron densities are in the range of N$_{e}$\,=\,70-400 cm$^{-3}$. In the northern and southern region of this galaxy, we found values ranging between N$_{e}$\,=\,10-700 cm$^{-3}$ and N$_{e}$\,=\,230-500 cm$^{-3}$ respectively. In six regions (dashed lines in Table~\ref{tableabundances_arp142}) we could not estimate electron densities because the sulphur emission-lines have low S/N.}
{The eastern regions present the lowest densities, with a mean electron density of N$_{e}$\,=\,300 cm$^{-3}$, below the critical value for collisional de-excitation (500 cm$^{-3}$). 
North-Western regions show the highest values of electron densities with a mean electron density of N$_{e}$\,=\,600 cm$^{-3}$ and the highest value was estimated in region S4A, where N$_{e}$\,=\,700 cm$^{-3}$.}

Finally, in a global view, our estimates of electron densities, which range from 10 cm$^{-3}$ to 700 cm$^{-3}$, are consistent with the values estimated by \citet{Krabbe:2014zi}  for a sample of local interacting galaxies. These authors found values ranging from 24 cm$^{-3}<$ Ne $<532$cm$^{-3}$, which are larger than the values obtained for a sample of isolated galaxies (40 cm$^{-3}<$Ne$<137$cm$^{-3}$). In this sense, the star-forming regions located in NGC\,2936 do not display the typical electron densities found in non-interacting systems. The origin of this trend is not clear. For example, \citet{Krabbe:2014zi} suggest that the high electron densities found in H{\sc ii} regions located in interacting galaxies do not seem to be produced by gas shocks.  However there is evidence from \citep[][e.g.]{Villa-Martin2014,Arribas2014} that does not allow us to exclude that high values detected near the nucleus are due to outflows  rather than shock-excited. Deeper studies on this topic are needed.
\subsubsection{Radial velocities}   

In the bottom right panel of Figure~\ref{Figure4} we show the radial velocity distribution (listed on Table \ref{tableradec}) of the star-forming regions observed in NGC\,2836. Radial velocities were derived by using the task {\sc emsao} in {\sc iraf}. The discrete data available does not allow us to study the whole kinematics of this system. It is clear from the figure that the southern tidal structure of NGC\,2936 displays lower radial velocities with respect to the main disk of this galaxy.  
By observing the H${\alpha}$ emissions in 2D spectra of the eastern region of NGC\,2936 ( the most intense spots of the left upper panel in Figure~\ref{Figure2}), we can see  that the kinematics of the region S1A  appears to be decoupled with respect to the kinematics of regions S1B and S1C with a difference in velocity of $\Delta v_r\sim 100 $kms$^{-1}$, which may be due to a distorted tidal arm in NGC\,2936. Integral field spectroscopy could be useful to disentangle the kinematics of region S1A.

\subsubsection{{H$\alpha$ velocity dispersion} }  
{{Despite the fact that the spectral resolution of our data is low (FWHM $\sim$207 km s$^{-1}$), estimating the velocity dispersion from the most prominent emission line, namely H$\alpha$, will give us a broader view of the internal kinematics of each region. Therefore, observed velocities dispersion ($\sigma_{H\alpha}$) were corrected for the instrumental velocity dispersion $\sigma_I$ and for thermal velocity dispersion $\sigma_T$ to obtain the intrinsic velocity dispersion $\sigma_{int}$,using the following equation: $\sigma_{int}=\sqrt{ {\sigma_{H\alpha}}^2  - {\sigma_I}^2 - {\sigma_T}}^2$. This exercise allow us to obtain the corrected velocities dispersion ($\sigma$(H$\alpha$)$_{int}$). Results are listed in the last column of Table \ref{tableabundances_arp142}. However, we note that our estimates should be read as referential values, considering the medium spectral resolution of our observations. Future high spectral resolution studies are needed to fully disentangle the internal kinematics of the star-forming regions in Arp 142.}}

\subsubsection{Ionization mechanism}

\begin{figure*}
\includegraphics[width=\columnwidth]{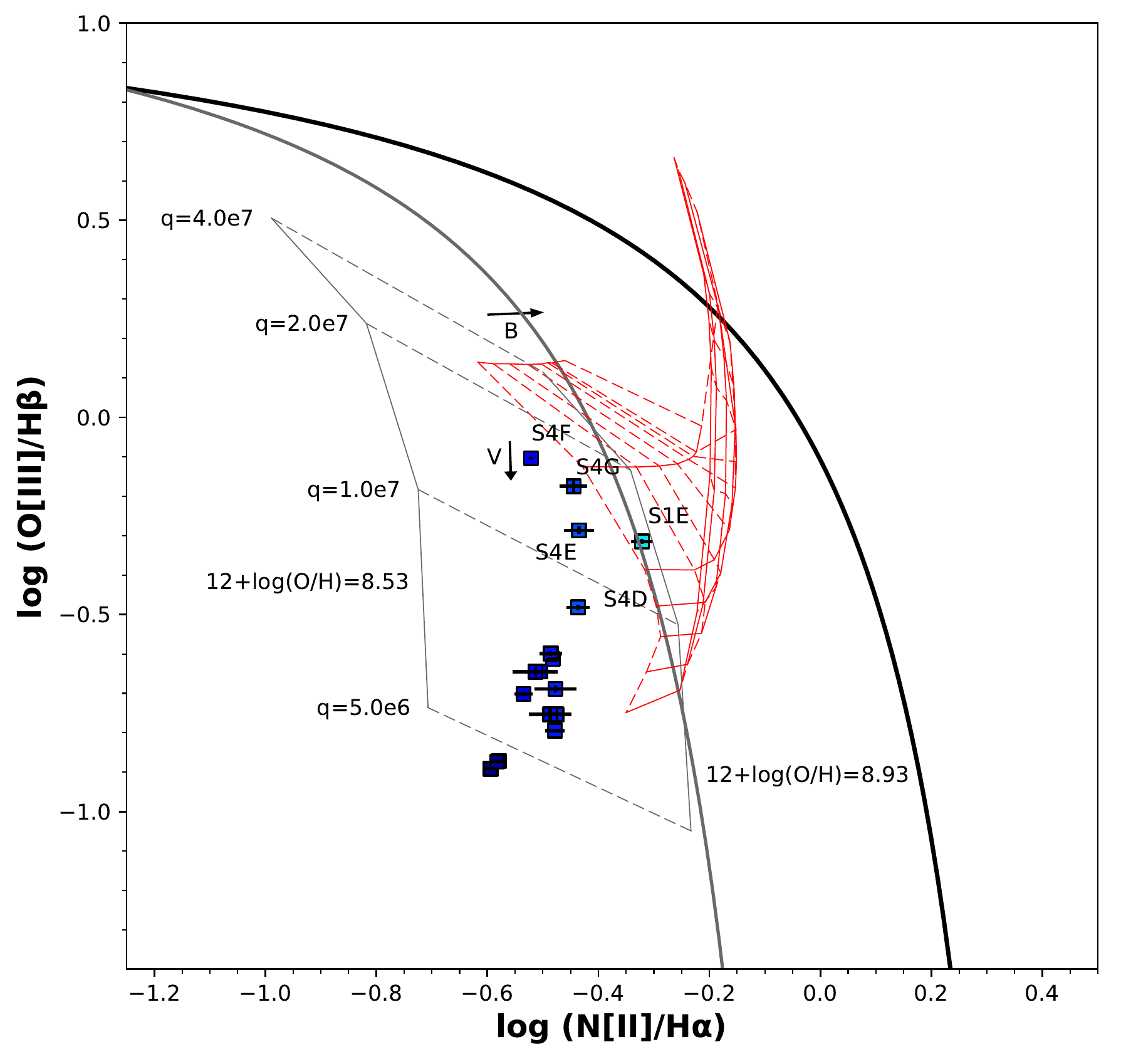}
\includegraphics[width=\columnwidth]{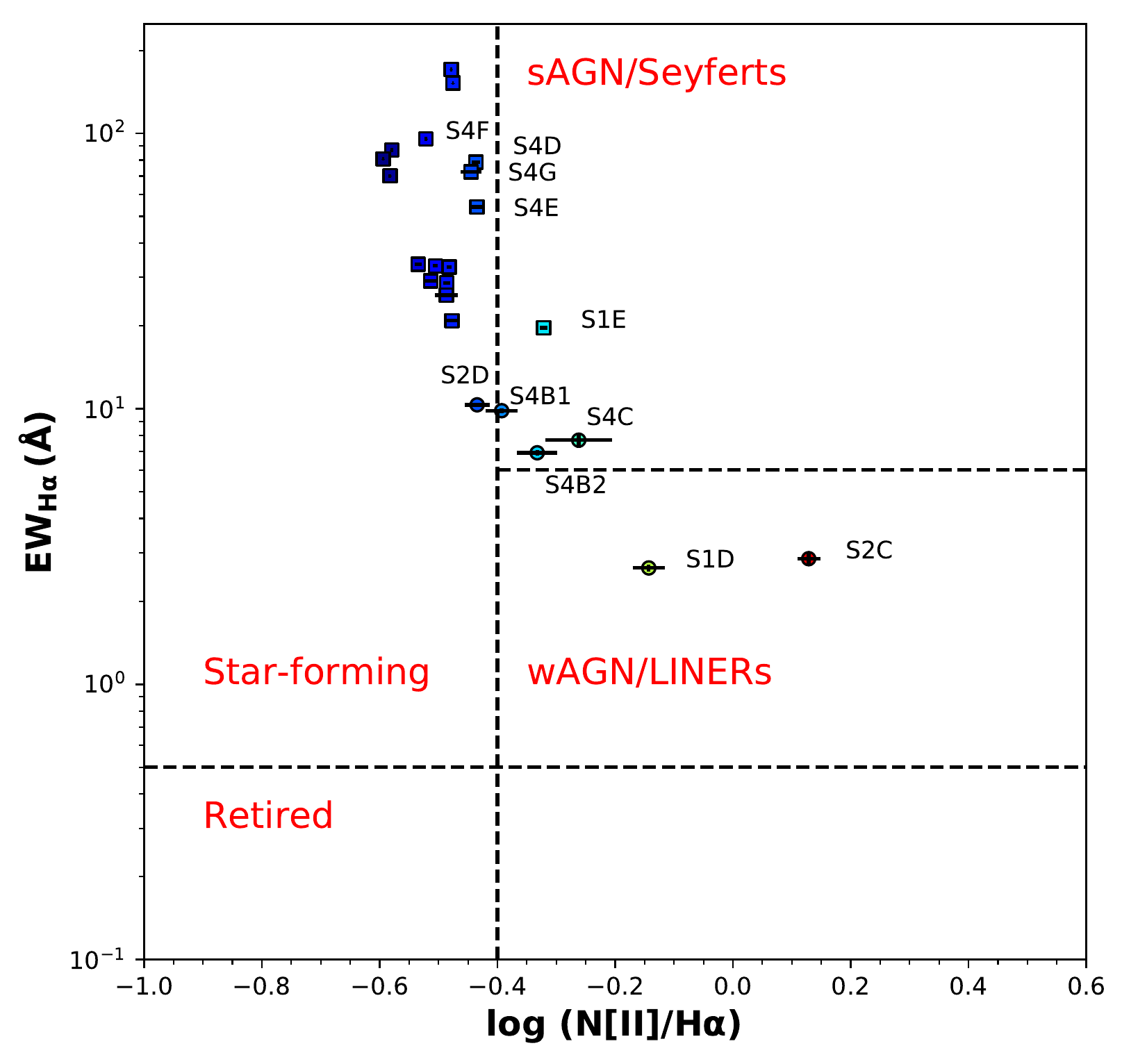}
\includegraphics[width=\columnwidth]{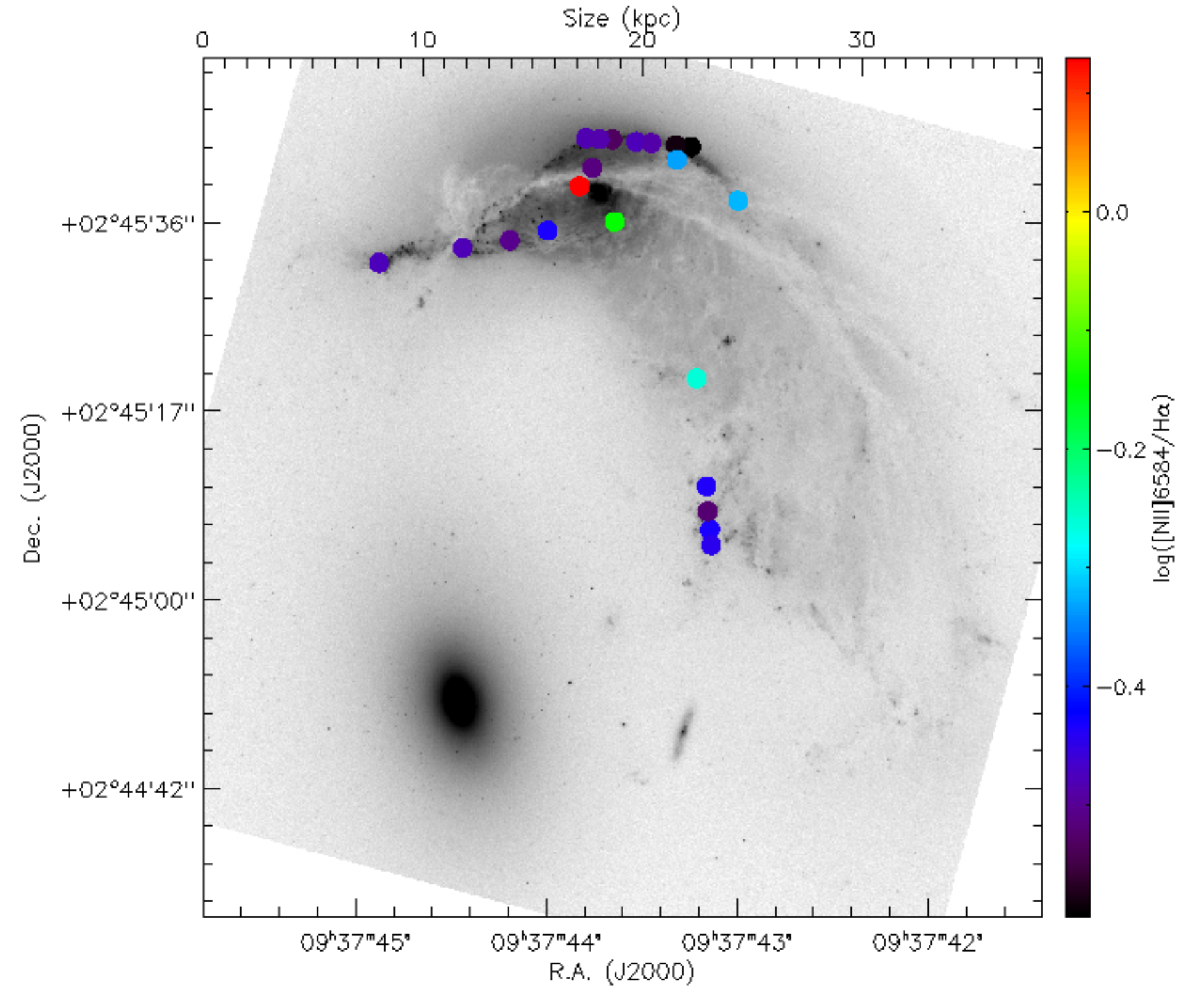} 

\caption{ {Diagnostic diagrams} to  identify the ionization mechanism for the regions observed in NGC\,2936. { Top-left: BPT diagram}. {{Continuous  {gray} line represents the limits suggested by \citet{Kauffmann:2003df} while {black} continuous line represents the limits from \citet{Kewley:2001ht}. Grey grid corresponds to photo-ionization model,  { while red grid correspond to shock-ionization models with a metallicity of
12+$\log$(O/H)=8.44} (see the text for more details of the models).  The color of the {squares}}} represent the [N{\sc ii}] over H$\alpha$ abundance distribution used for color coding on the bottom panel. {The arrows near B and V letters indicate the
direction in which the magnetic field
and shock velocity increment, respectively, in the models.} { Top-right: WHAN diagram.
The circles symbols come from nuclear and plume regions
and are not present in
BPT diagram. The color-coded is the same as BPT diagram.
The dashed vertical line at [N{\sc ii}]/H$\alpha=-0.4$ represents the optimal transposition between star-forming and nuclear activity (or shocks).
The horizontal line at EW$_{H\alpha}=6$ represents
the optimal transposition between Seyferts and LINERs,
while the one at EW$_{H\alpha}=0.5$ represents the
division for retired galaxies \citep{2011MNRAS.413.1687C}}. Bottom: Color-coded spatial distribution of the regions by the [N{\sc ii}] over H$\alpha$ abundances.   {Figure clearly illustrated that both the nuclear and plume regions are being ionizing 
by AGN/Shock activities}.}
\label{Figure5}
\end{figure*}

Most of the sources for which we have spectroscopic information resemble star-forming objects. However, H{\sc II} regions can be  either photo-ionized  or shock-ionized \citep[e.g.][]{2004AJ....127.1405C,2011hong,2013hong, 2015ApJS..221...28R,jose}. In order to determine and confirm the ionization mechanism of the different regions of NGC\,2936 we have plotted their [N{\sc ii}]/H$\alpha$ versus [O{\sc iii}]/H$\beta$ line ratio in a diagnostic Baldwin-Phillips-Terlevich (BPT) diagram \citep[][]{Baldwin:1981qm}, 
which is shown in the top panel of Figure~\ref{Figure5}. On the BPT diagnostic diagram, 
{{we have included a grid of  photoionization models calculated by \citet{Kewley:2001ht}. These models are based on stellar population synthesis models STARBURST99 \citep{1999ApJS..123....3L}
in conjunction with a photoionization code Mappings III \citep{1985ApJ...297..476B, 1993ApJS...88..253S}.
STARBUST99 generates the spectral energy distribution of a stellar cluster that serves as an ionization source that is used  as input for MAPPING III calculating the ionization states and the  emission line fluxes of the elements of the gas \citep[for further details see][]{Kewley:2001ht}. The photoionization model grid (in gray color) shown in the BPT of  the Figure~\ref{Figure5} diagram has two metallicities (plotted as continuous lines): 12+$\log$(O/H)=8.53 and 12+$\log$(O/H)=8.93, covering the range of the measured metallicities listed in Table \ref{tableabundances_arp142}), and the ionization parameter (q) ranges from $5.0\times10^6$ up to $4.0\times10^7$\,cm\,s$^{-1}$ (dashed lines).
These models are calculated assuming an instantaneous starburst and a gas density of 10\,cm$^{-3}$.  As a reference, we also plotted  the photoionized maximum limits for H{\sc ii} proposed by \citet{Kauffmann:2003df} (empirically) and  \citet{Kewley:2001ht} (theoretical), which are shown in grey and  black lines respectively.  In addition to the photo-ionization models, we plotted a grid of shock-ionization models from  \citet{2008ApJS..178...20A}. These authors use the Mappings III shock
and photo-ionization code to calculate the ionization states and the line emission fluxes of the gas. The ionizing radiation field input is produced by the cooling zone behind the front shock (post-shock region) and it is mostly composed of thermal bremsstrahlung (free-free) continuum. The main parameters of the shock models are: the shock velocity, magnetic field (B), pre-shock gas density, and metallicity. The shock emission of the models can be divided into two components: the radiative shock (post-shock region) and its photo-ionized precursor (pre-shock region). In the BPT diagrams, we plot { a grid} with only the 
radiative shock for two different  metallicities:  12+$\log$(O/H)=8.44 (in red) and 12+$\log$(O/H)=8.93 (in blue), both with the same 
electronic density of 1\,cm$^{-3}$. The shock velocity (continuous lines)
range is  100\,$-$\,700 km\,s$^{-1}$ and 200\,$-$\,1000 km\,s$^{-1}$, respectively.
The ISM magnetic field (dashed lines) range is $1\times10^{-4}\,-\,$10\,$\mu$G.}  {Because of the huge extinction  of the nuclear and plume regions, H$\beta$ line is not present on the aperture spectra; hence its corresponding points are not on BPT diagram. In order to overcome this, we used the WHAN diagram \citet{2011MNRAS.413.1687C}: the H$\alpha$ equivalent width (EW$_{H\alpha}$) vs [N{\sc ii}]/H$\alpha$. In this plot all sources are present (Top-left panel Figure
~\ref{Figure5}). The dash vertical line at [N{\sc ii}]/H$\alpha=-0.4$ separates the star-forming and AGN galaxies. The Star-forming regions falling into the AGN sector could be interpreted as being ionized by shocks.}}
On the bottom panel of Figure~\ref{Figure5} we display an optical \textit{HST} image of Arp\,142, where the circles represent the location of the different spectroscopic sources.

Inspecting the {{top-left}} panel of Figure~\ref{Figure5} we  found that {all regions lying on the spiral arms (dark blue and black squares) fall on the quadrant corresponding to the lower ionizing parameter of the photo-ionization models (considering the uncertainties).
The regions associated to the {{ plume (S4D, S4E, S4F, S4G, and S1E)}) are located inside  a ``mixing'' zone; they fall  near or inside the quadrant of the lower shock velocity (i.e. 200\,$-$\,300 km\,s$^{-1}$)  of the shock-ionization model 
with a metallicity of  12+$\log$(O/H)=8.44. They are also inside the quadrant of q=1.0e7 and q=2.0e7 of the photo-ionization model. Despite this degeneracy,  it is more probable that these regions are shock-ionized rather that photo-ionized because they are close to the line representing a metallicity of 12+$\log$(O/H)=8.93 in the  photo-ionization model, which is nearly
0.3\,dex higher than metallicities found for these regions. 
{In addition,  the other sources present in the plume (S4B1, S4B2 and S4C), but no H$\beta$, falling into the AGN sector in the WHAN diagram (top-right panel of Figure \ref{Figure5}) indicate that they are also being ionized by shocks.} Moreover in the regions located at the plume, the tidal effect is more pronounced (due to they are at the outskirt of the galaxy), and therefore, a shock excitation is  more likely.  {{Finally, from the kinematic point of view, their  velocity dispersion measured are super-sonic (see Table \ref{tableabundances_arp142}), therefore, reinforcing the above shock scenario.}}
{ The locus of regions close to the nucleus (S2C and S1D) 
in WHAN diagram lie on the quadrant of AGN activity, suggesting 
that the ionization of the gas is dominated by nuclear activity
and/or shocks.}This scenario is consistent with the \textit{SDSS} nuclear classification of NGC\,2936, which is cataloged as an AGN galaxy.} }


\begin{table*}
\begin{minipage}[t]{\textwidth}
\centering
\scriptsize
\caption{Oxygen abundances and star formation rates (corrected and uncorrected from internal extinction).}
\begin{tabular}{cccccccccccc}
\hline
ID  & n([S{\sc ii}])& L$_{H\alpha}$uncor & E(B-V) & L$_{H\alpha}$cor  &  12+log(O/H)$_{O_{3}N_{2}}$  & SFR$_{H\alpha}$uncor     & SFR$_{H\alpha}$cor   & Area      & log $\Sigma_{SFR,cor}$ &  FWHM(H$\alpha$) & $\sigma$(H$\alpha$)$_{int}$ \\
     & cm$^{-3}$ &  $\times10^{39}$ erg s$^{-1}$       & mag     &  $\times10^{40}$ erg s$^{-1}$           &          &    M$_{\odot}$ yr$^{-1}$       & M$_{\odot}$ yr$^{-1}$  & kpc$^{2}$ & M$_{\odot}$ yr$^{-1}$ kpc$^{-2}$ &\AA & kms$^{-1}$\\
\hline
S1A  &	70 $\pm$ 10		    	&	163.0	$\pm$	11.7	&	 0.30 $\pm$	$<$0.01& 39.8	$\pm$	2.8	&	8.59	$\pm$	0.18	&	1.28	$\pm$	0.09		&	 3.15	$\pm$	0.22		&	1.00 $\pm$ 0.04	&	 0.50 $\pm$ 0.05  	&4.48 $\pm$ $<$0.01	& ---\\
S1B	&	400 $\pm$ 50			&	205.0	$\pm$	18.1	&	 0.80 $\pm$	0.11&	226.0	$\pm$	16.2	&	8.60	$\pm$	0.18	&	1.62	$\pm$	0.14		&	17.88	$\pm$	1.28		&	0.90 $\pm$ 0.03	&	 1.30 $\pm$ 0.05	&5.33 $\pm$  0.01 	& 41 $\pm$ 9\\
S1C	&	200 $\pm$ 70 		&	44.4 	$\pm$	3.5	&	 0.49 $\pm$	0.05&	19.2 	$\pm$	1.4	&	8.56	$\pm$	0.18	&	0.35	$\pm$	0.03		&	 1.52	$\pm$	0.11		&	0.90 $\pm$ 0.03	&	 0.23 $\pm$ 0.05	& 4.57 $\pm$ 0.01 	& ---\\
S1D$^{\circ}$	& - - -			&	7.6		$\pm$	0.8	&	  ---           		        &	0.8		$\pm$	0.1	&	---                  		&	0.06	$\pm$	0.01		&	 0.06	$\pm$	0.01		&	2.40 $\pm$ 0.09	&	-1.60 $\pm$ 0.09	& 4.96 $\pm$ 0.13 	& 14 $\pm$ 10 \\
S1E	&	 50:$^{\S}$ 			&	10.8		$\pm$	1.2	&	 0.58 $\pm$	0.14&	6.2		$\pm$	0.5	&	8.53	$\pm$	0.18	&	0.09	$\pm$	0.01		&	 0.49	$\pm$	0.04		&	1.10 $\pm$ 0.04	&	-0.35 $\pm$ 0.05	&6.09 $\pm$  0.04  & 71 $\pm$ 8 \\
S2A	&	180:$^{\S}$			&	54.4		$\pm$	4.1	&	 0.24 $\pm$	0.01&	11.1		$\pm$	0.8	&	8.57	$\pm$	0.18	&	0.43	$\pm$	0.03		&	 0.87	$\pm$	0.06		&	0.50 $\pm$ 0.02	&	 0.24 $\pm$ 0.05	& 6.17 $\pm$ 0.02 	& 74 $\pm$ 8\\
S2B	&	200:$^{\S}$ 			&	30.6		$\pm$	2.4	&	 0.07 $\pm$	$<$0.01&3.7		$\pm$	0.3	&	8.56	$\pm$	0.18	&	0.24	$\pm$	0.02		&	 0.30	$\pm$	0.02		&	0.30 $\pm$ 0.01	&	-0.01 $\pm$ 0.04	& 6.15 $\pm$ 0.04 	& 73 $\pm$ 8 \\
S2C$^{\circ}$	&	- - -			&	18.3		$\pm$	2.0	&	 ---            		        &	1.8		$\pm$	0.2	&	---                  		&	0.14	$\pm$	0.02		&	 0.14	$\pm$	0.02		&	0.70 $\pm$ 0.03	&	-0.68 $\pm$ 0.08	& 7.71 $\pm$ 0.31 	& 117 $\pm$ 9\\
S2D	&	- - -					&	7.4		$\pm$	1.7	&	 0.78 $\pm$	0.61&	7.5		$\pm$	0.6	&	---                 		&	0.06	$\pm$	0.01		&	 0.59	$\pm$	0.05		&	0.40 $\pm$ 0.01	&	 0.17 $\pm$ 0.05	& 5.54 $\pm$ 0.11 	& 51 $\pm$ 9\\
S3A	&	10:$^{\S}$			&	55.1		$\pm$	4.4	&	 0.31 $\pm$	0.01&	13.7		$\pm$	1.0	&	8.56	$\pm$	0.18	&	0.44	$\pm$	0.03		&	 1.08	$\pm$	0.08		&	0.40 $\pm$ 0.01	&	 0.43 $\pm$ 0.05	& 5.90 $\pm$ 0.02 	& 65 $\pm$ 9 \\
S3B	&	10:$^{\S}$			&	19.0		$\pm$	1.5	&	 0.20 $\pm$	0.02&	3.5		$\pm$	0.3	&	8.56	$\pm$	0.18	&	0.15	$\pm$	0.01		&	 0.27	$\pm$	0.02		&	0.20 $\pm$ 0.01	&	 0.13 $\pm$ 0.05	& 5.44 $\pm$ 0.01 	& 46 $\pm$ 9\\
S3C	&	30:$^{\S}$ 			&	22.0		$\pm$	1.7	&	 0.00 $\pm$	0.01&	2.2		$\pm$	0.2	&	8.58	$\pm$	0.18	&	0.17	$\pm$	0.01		&	 0.17	$\pm$	0.01		&	0.20 $\pm$ 0.01	&	-0.06 $\pm$ 0.04	&  5.20$\pm$ 0.03	& 34 $\pm$ 9\\
S3D	&	60:$^{\S}$			&	9.6		$\pm$	0.7	&	 0.00 $\pm$	$<$0.01&1.0		$\pm$	0.1	&	8.59	$\pm$	0.18	&	0.08	$\pm$	0.01		&	 0.08	$\pm$	0.01		&	0.20 $\pm$ 0.01	&	-0.42 $\pm$ 0.07	& 5.29 $\pm$ 0.03	& 39 $\pm$ 9 \\
S3E$^{\dag}$ &500 $\pm$ 90	&	89.9		$\pm$	6.9	&	 0.54 $\pm$	0.02&	44.3		$\pm$	3.2	&	8.60	$\pm$	0.18	&	0.71	$\pm$	0.05		&	 3.50	$\pm$	0.25		&	0.40 $\pm$ 0.01	&	 0.94 $\pm$ 0.05	&  4.61$\pm$ 0.01 	& ---\\
S3F	&	600 $\pm$ 50			&	55.7		$\pm$	4.2	&	 0.53 $\pm$	0.02&	26.9		$\pm$	1.9	&	8.60	$\pm$	0.18	&	0.44	$\pm$	0.03		&	 2.12	$\pm$	0.15		&	0.30 $\pm$ 0.01	&	 0.85 $\pm$ 0.05	& 4.47 $\pm$ 0.01 	& ---\\
S4A$^{\dag}$& 700:$^{\S}$		&	94.1		$\pm$	7.0	&	 0.43 $\pm$	0.01&	33.7		$\pm$	2.4	&	8.60	$\pm$	0.18	&	0.74	$\pm$	0.06		&	 2.66	$\pm$	0.19		&	1.40 $\pm$ 0.05	&	 0.28 $\pm$ 0.05	& 5.15 $\pm$ 0.01 	& 31 $\pm$ 9\\
S4B1$^{\circ}$	&	- - -	 	&	2.4		$\pm$	0.2	&	 ---            		         &	0.2		$\pm$	0.02	&	---                 		&	0.02 $\pm$	$<$0.01	&	 0.02	$\pm$	$<$0.01	&	0.50 $\pm$ 0.02	&	-1.43 $\pm$ 0.06	& 4.60 $\pm$ 0.10 	& ---\\
S4B2$^{\circ}$	& - - -		&	1.2		$\pm$	0.1	&	 ---            		         &	0.1		$\pm$	0.01	&	---                  		&	0.01	$\pm$	$<$0.01	&	 0.01	$\pm$	$<$0.01	&	0.30 $\pm$ 0.01	&	-1.50 $\pm$ 0.06	& 4.45 $\pm$ 0.10 	& --- \\
S4C	& - - -					&	1.5		$\pm$	0.4	&	 0.36 $\pm$	0.37&	0.4		$\pm$	0.1	&	---                	    	&	0.01	$\pm$     $<$0.01	&	 0.04	$\pm$	$<$0.01   &	0.30 $\pm$ 0.01	&	-0.93 $\pm$ 0.05	& 5.53 $\pm$ 0.37 	& 50 $\pm$ 10 \\
S4D	&	300:$^{\S}$			&	7.6		$\pm$	0.8	&	 0.64 $\pm$	0.14&	5.1		$\pm$	0.4	&	8.54	$\pm$	0.18	&	0.06	$\pm$	0.01		&	 0.40	$\pm$	0.03		&	0.70 $\pm$ 0.03	&	-0.24 $\pm$ 0.05	& 4.56 $\pm$ 0.03 	& ---\\
S4E	&	230:$^{\S}$			&	5.0		$\pm$	0.4	&	 0.22 $\pm$	0.04&	1.0		$\pm$	0.1	&	8.50	$\pm$	0.18	&	0.04	$\pm$	$<$0.001	&	 0.08	$\pm$	0.01		&	0.70 $\pm$ 0.03	&	-0.97 $\pm$ 0.07	& 5.49 $\pm$ 0.04 	& 49 $\pm$ 9\\
S4F	&	400 $\pm$ 50			&	24.5		$\pm$	1.8	&	 0.24 $\pm$	0.01&	4.9		$\pm$	0.4	&	8.44	$\pm$	0.18	&	0.19	$\pm$	0.01		&	 0.39	$\pm$	0.03		&	1.20 $\pm$ 0.04	&	-0.49 $\pm$ 0.05	& 4.99 $\pm$ 0.01 	&19 $\pm$ 9\\
S4G	&	500:$^{\S}$			&	2.8		$\pm$	0.3	&	 0.26 $\pm$	0.07&	0.6		$\pm$	0.05	&	8.48	$\pm$	0.18	&	0.02	$\pm$	$<$0.001	&	 0.05	$\pm$	$<$0.001	&	0.30 $\pm$ 0.01	&	-0.80 $\pm$ 0.05	& 7.18 $\pm$ 0.02 	& 103 $\pm$ 8\\
\hline
\multicolumn{10}{@{}l@{}}{\footnotesize $^{\dag}$Same star-forming region observed in two different slits (S3 and S4).}\\
\multicolumn{10}{@{}l@{}}{\footnotesize $^{\circ}$ Star-forming region uncorrected by extinction.}\\
\multicolumn{10}{@{}l@{}}{\footnotesize $^{\S}$Due to the very large uncertainties ($\geq$50\%), these density values are only estimates of the order of magnitude of the electron densities.} \\
\multicolumn{10}{@{}l@{}}{\footnotesize Dash lines mean sulphur emission-lines are very noisy and not density values were estimated.}\\
\multicolumn{10}{@{}l@{}}{\footnotesize$\sigma$(H$\alpha$)$_{int}$: Intrinsic velocity dispersion corrected by instrumental and thermal dispersion.} \\
\label{tableabundances_arp142}
\end{tabular}
\end{minipage}
\end{table*}


\subsubsection{The color image of the Arp\,142 system}
\label{colorimage}
\begin{figure}
\includegraphics[width=\columnwidth]{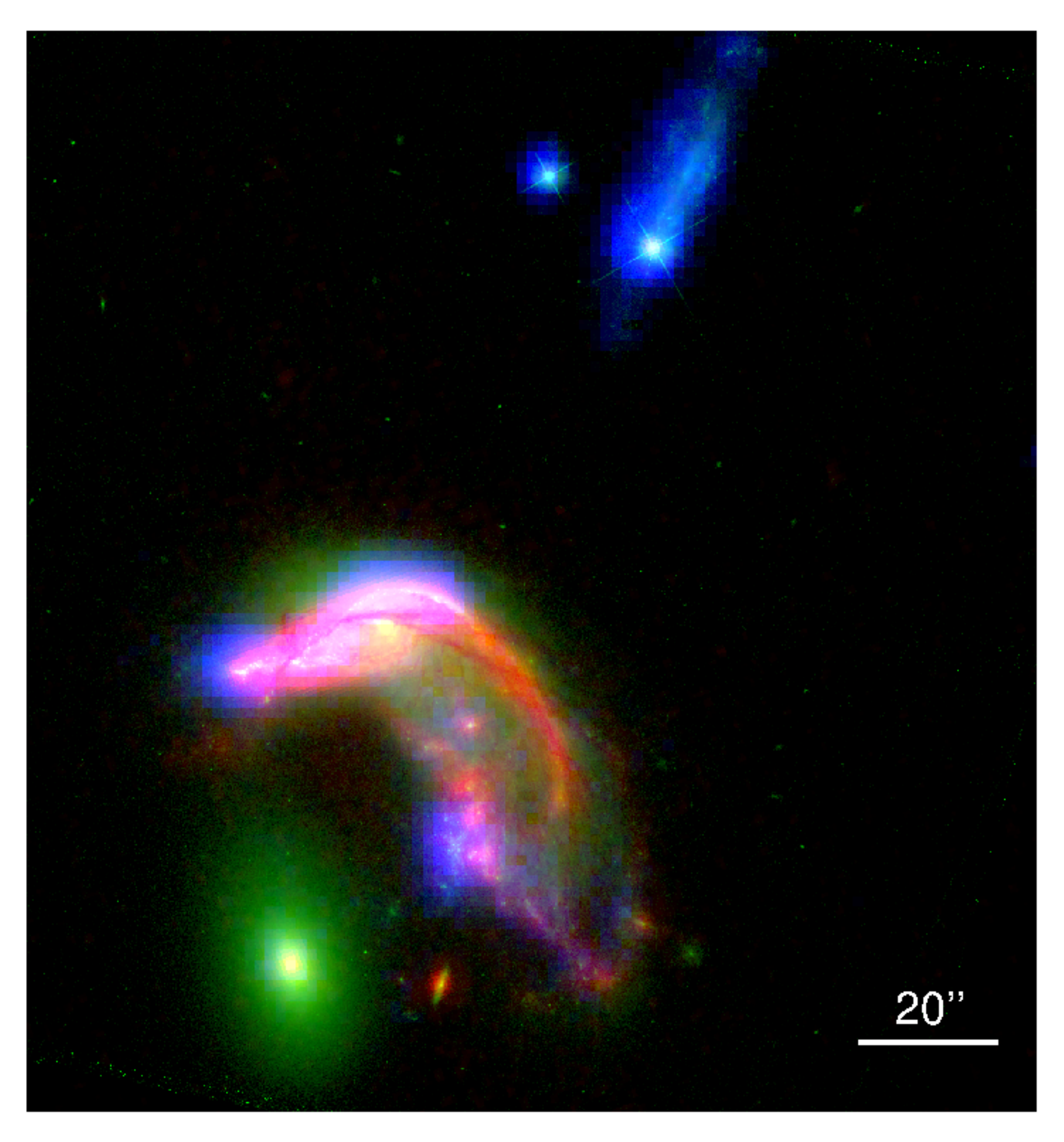}
\caption{RGB Color composite figure of Arp\,142. Blue color corresponds to the NUV from Galex, Green to the HST $F606W$ and Red to Spitzer 8 microns.}
\label{Figure6}
\end{figure}

{
In Figure~\ref{Figure6} we show a color image of Arp\,142, which is a composition of NUV/\textit{GALEX} (blue),  $F606W$/\textit{HST} (green) and 8$\mu$m/\textit{Spitzer} images. It is clear from this image that the very strong dust lane seen on the \textit{HST} image is evidenced by \textit{Spitzer} at 8$\mu$m.The blue color maps the young star-forming regions, and it is in agreement with the top left panel of the Figure \ref{Figure4}, while the magenta color maps the embedded/newly formed star-forming regions. In order to quantify the star formation rate on this system by using archival data, we have used NUV/\textit{GALEX} and \textit{Spitzer} already published in the literature. In the case of NUV/\textit{GALEX}, we have used the magnitude published by \citet{2007ApJS..173..185G} and the equation (3) in \citet{2006ApJS..164...38I}. This exercise provides a SFR$_{NUV}$=3.8 M$_{\odot}$ yr$^{-1}$ once the NUV emission has been corrected by dust (equation 1 from that paper). If we do not consider the dust correction, the star formation rate drop to SFR$_{NUV}$=0.9 M$_{\odot}$ yr$^{-1}$ (using equation 3 in \cite{2006ApJS..164...38I}). In the case of the IR information, we can use it to trace the star formation rate associated with dust heated by energetic photons. By using the \textit{total} IR emission in NGC\,2936 (listed in Table 3 of \citet{Xu:2010ix}) and the equation 5 of \citet{2006ApJS..164...38I}, we derived a SFR$_{total IR}$=12.6 M$_{\odot}$ yr$^{-1}$. This value is slightly lower if we consider the corrected IR luminosity displayed in Table 3 of \citet{Xu:2010ix}. Finally, H$\alpha$ emission lines from the star-forming regions  were also used to get star formation rates and added up to give a number for the whole system. The latter is,  of course, a lower limit for the total star formation rate of the system. The three star formation rate estimations obtained from NUV/\textit{GALEX} (corrected by dust), \textit{Spitzer} (total IR emission published by \citet{Xu:2010ix}) and H$\alpha$ (once corrected by internal extinction)  display different results (3.8 M$_{\odot}$ yr$^{-1}$ with \textit{GALEX}, 12.6 M$_{\odot}$ yr$^{-1}$ with \textit{Spitzer} and 35.9 M$_{\odot}$ yr$^{-1}$ for H$\alpha$). In this context, \citet{2006ApJS..164...38I} found that the star formation rates traced by IR emission can exceed the values derived from the UV emission (by a factor 2), where this discrepancy could arise from the dust attenuation correction for dusty galaxies. {In the case of the SFRs derived from H$\alpha$, we note that dust correction plays an important role. As previously mentioned, the \citet{Fitzpatrick:1999ca} extinction law allowed us to derive a lower SFR for region S1B (8.7 M$_{\odot}$ yr$^{-1}$ versus 17.9 M$_{\odot}$ yr$^{-1}$ once Calzetti extinction law was used). If we scale this difference to the total SFR in NGC 2936 (derived from H$\alpha$), we found a total value of 17.5 M$_{\odot}$ yr$^{-1}$, which is not so different with respect to the SFR derived from IR information (SFR$_{totalIR}$=12.6 M$_{\odot}$ yr$^{-1}$). In this sense, a precise determination of the dust correction is mandatory to compare the different SFR tracers.} In any case, the values that we obtain for the spiral member in Arp\,142 show an actively star-forming object, where the interaction with NGC\,2937 should be playing an important role. Finally, we note that \citet{Xu:2010ix} has also obtained star formation rates from \textit{Spitzer}, and we (not surprisingly) get the same result, given that we also used the same data.}

\subsection{The velocity of the galaxy north of Arp\,142}

The elliptical galaxy NGC\,2937 is only 0.84 arcmin or 24.4 kpc projected distance from NGC\,2936. {{Therefore it is the natural candidate responsible for the distorted companion spiral galaxy.}} However, there is also another possibility, which is a small galaxy north of the system, UGC\,05138\,NOTES01, at a projected distance of 1.3 arcmin north of the center of NGC\,2936. We have placed one of the slits on this galaxy and its spectrum is then shown in Figure. \ref{Figure3}, bottom panel. From the several emission-lines, we could obtain a radial velocity for this galaxy of 4971$\pm$18 km s$^{-1}$, which is $\sim$ 2000 km s$^{-1}$ different from the Arp\,142 system. Therefore we conclude that this galaxy cannot be the disturbing agent that caused the morphological and kinematic changes of NGC\,2936.

\subsection{{Faint tidal bridge between NGC\,2936 and NGC\,2937}}
\label{bridge}

\begin{figure}
\includegraphics[width=\columnwidth]{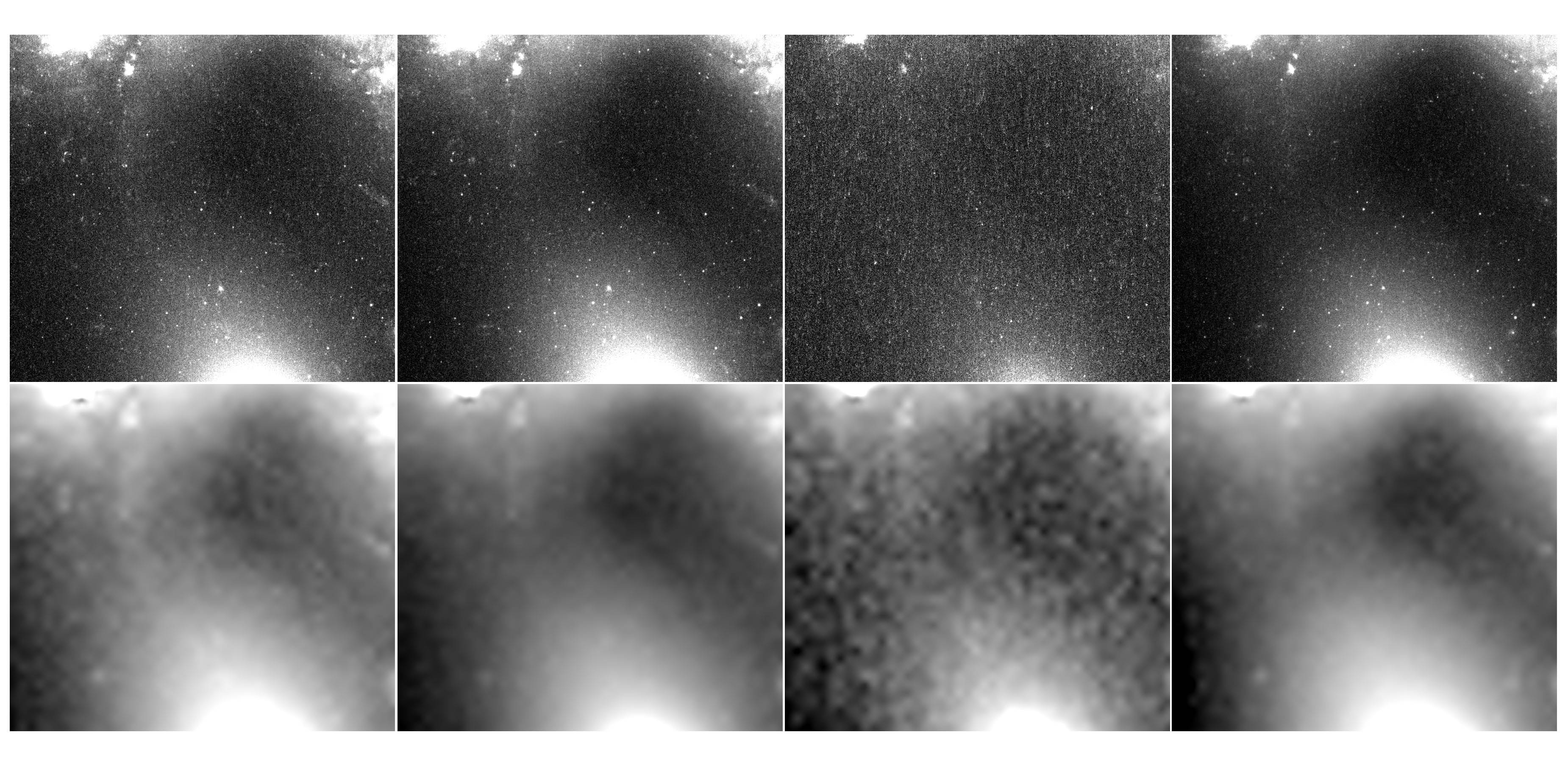}
\caption{Tidal feature derived from HST image. From Right to left Each stamps corresponds to 36" x 32" Top stamps: From left to right: $F475W$, $F606W$, $F657N$ and, $F814W$ . Bottom: TPS modeled images. From left to right: $F475W$, $F606W$, $F657N$ and, $F814W$}
\label{Figure7}
\end{figure}

{Low surface brightness merger indicators (i.e. faint tidal tails, bridges, etc) are often missed on images due to the lack of sufficient sensitivity or short exposure times. In particular, when we were selecting the interesting regions to be pointed by GMOS, we noted that there is a faint tidal arm that appears to originate in the center of the elliptical galaxy NGC\,2937 that runs in the direction of the spiral galaxy NGC\,2936. This can be seen by eye but it is hard to display it in a figure, given that it is a very faint feature. In order to evidence this feature and to eventually highlight other faint and extended features in the image, we mapped the background (i.e. the faint and extended component) using a Thin Plate Spline Interpolation (TPS). Briefly, TPS is a 2D interpolation for arbitrarily spaced data forming a grid. The TPS analogy is to consider a thin metal sheet that is shaped by a grid, forcing the sheet to not move at the points that are modeled on the grid. By applying this technique, the feature we had noted by eye could then be revealed (see the bottom row of Figure~\ref{Figure7}). Since this technique is computationally expensive, we focused on the region in between the galaxies. We applied the TPS considering a grid separated by 20 x 20 pixels, which is a size large enough to ignore the stars but small enough to map extended sources. The output is a new image with the background modeled and that contains the extended sources of the image.  In the upper row of Figure~\ref{Figure7} the original HST images are shown and  results from the TPS are shown in the bottom row of Figure~\ref{Figure7} where the tidal {bridge} is clearly seen on the bluest HST filter ($F475W$) and it is less clearly seen in the red one $F814W$. The narrow-band $F657N$ frame does not clearly show {the tidal bridge}, probably due to the shallower image, compared with those for the other filters.}

\subsection{Isophotal Analysis of NGC\,2937}
\label{sec:iso}

{Unlike to the spiral galaxies, which are ``cold'' systems, the ellipticals response
to tidal fields in a subtle fashion because they are ``hot'' systems. The most striking tidal structures
observed in ellipticals are the shells and ripples, however, they are relics/debris of a past interaction rather than tidal structures of an ongoing event \citep{Barnes1992}. The latter events can produce effects such as enhancement in surface brightness profile 
at outer radii \citep{Kormendy3_1977},   non-centric inner isophotes 
\citep[e.g.][]{Lauer1986,Lauer1988,Gonzales-Serrano2000}, and  ``u-shape'' velocity profiles \citep[e.g.][]{Borne1988}.  
The first two of these features we are able to probe by performing a photometry analysis on $F475W$ (SDSS's $g$-filter) HST image of NGC\,2937. The magnitude zero point (ZP) of the image was obtained from the ACS webpage corresponding to the date of the observation, for  ZP$_{(F475W)}$ = 25.75.
}

\begin{figure}
\includegraphics[width=\columnwidth]{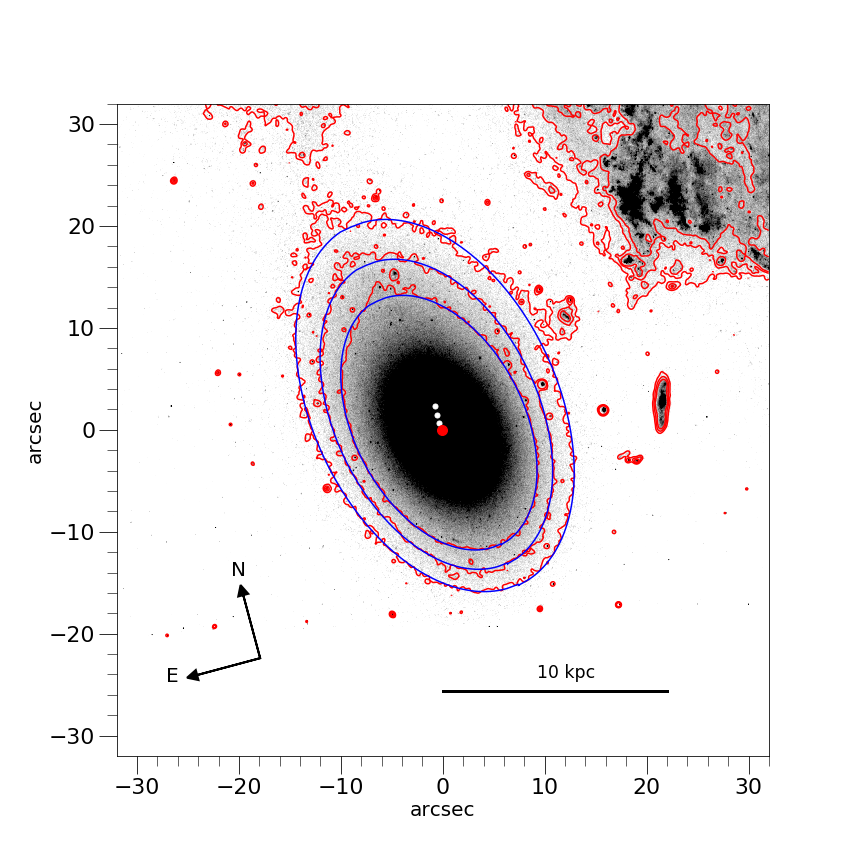}
\includegraphics[width=\columnwidth]{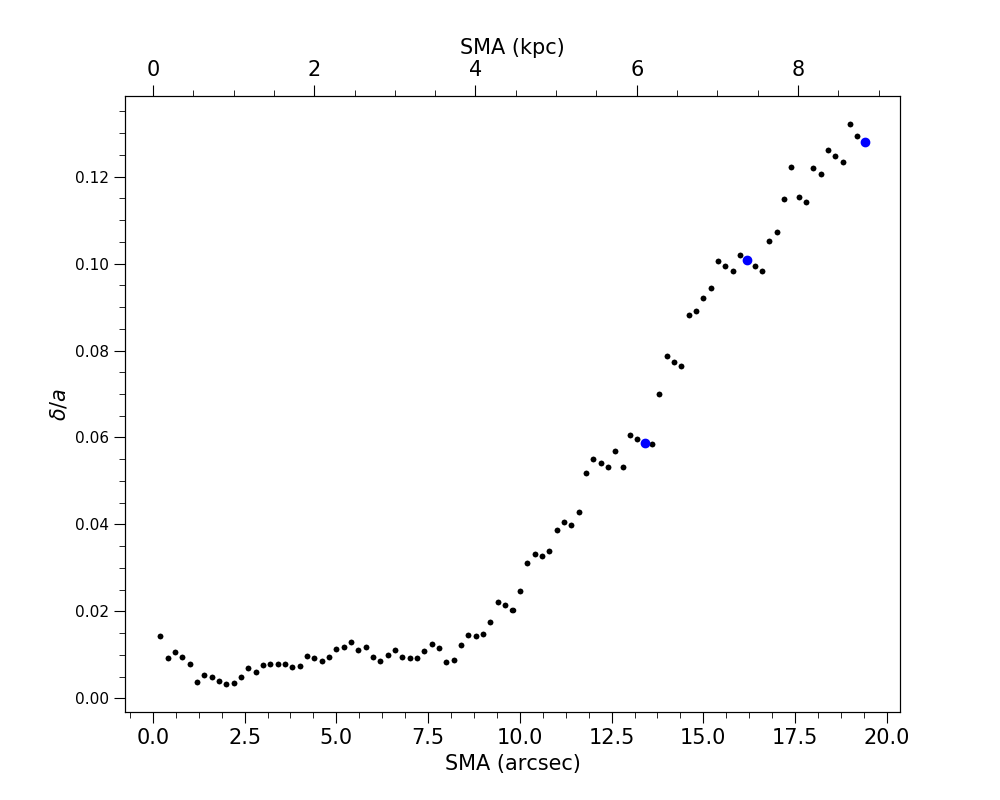}
\includegraphics[width=\columnwidth]{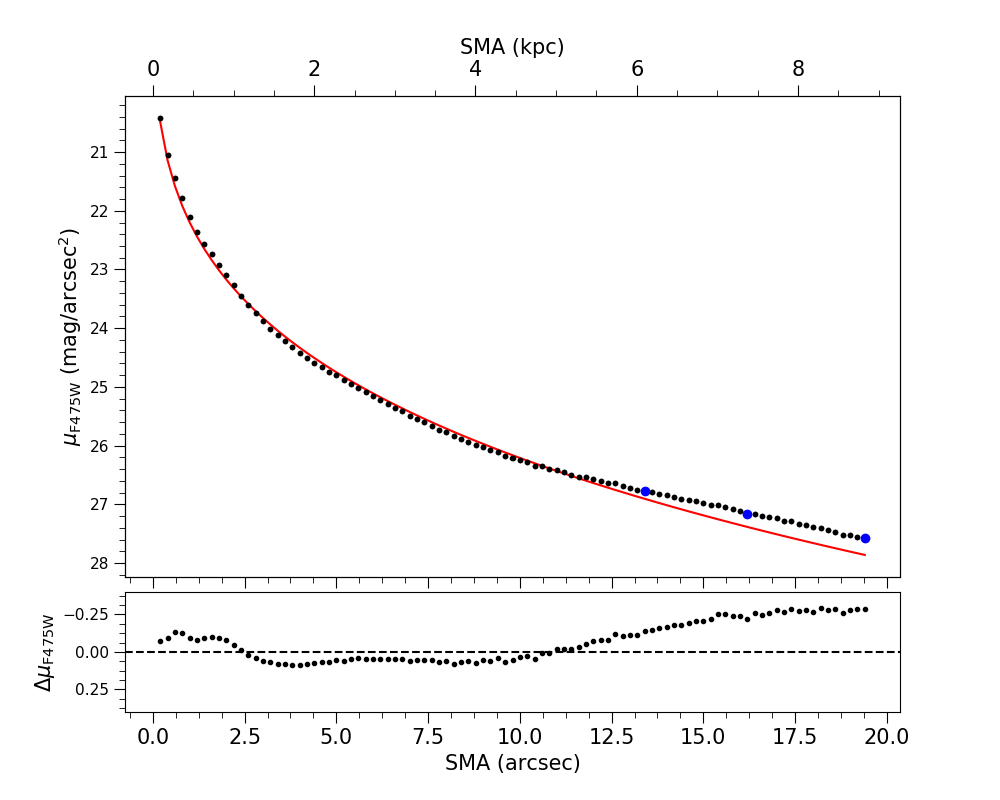}
\caption{Top panel: Selected fitted ellipses (shown in blue) that overlaid  the isophote contours (shown in red). The {{ white circle}} markers correspond to the ellipse centers, while the red {one} is the photometric center of the galaxy. Middle panel: The $\delta$/a ratio profile. Bottom panel: $F475W$ band surface bright profile (black dot-points) fitted  with de Vaucouleurs law (red solid line). The small box below the figure corresponds to the residuals {{in mag/arcsec$^2$.  The blue points in middle and bottom panels correspond to the selected isophotes showed in top panel.}}}
\label{Figure8}
\end{figure}

We have modeled the isophotes of NGC\,2937 by using the {\sc iraf} task $ellipse$. The geometrical parameters; position angle, ellipticity, and center of the ellipses were let free. To avoid the substructures or bad pixels the clipping algorithm was activated with an  allowed maximum fraction of 20\% of flagged pixels.
{{Top panel of}} Figure~\ref{Figure8} shows {{some selected fitted ellipses on outer isophotes}}. It can be clearly seen that the centers of {{these}} outermost isophotes are offset toward NGC\,2936 at the same direction as the faint tidal feature  found out in Figure~\ref{Figure7} (see Section 3.4). One way to quantify this displacement is by the $\delta/a$ ratio \citep{Lauer1988},  where $\delta$ is the separation of isophote center with respect to the nucleus and $a$ is the length of Semi-Major Axis (SMA). The radial profile of this ratio is shown
in the middle panel of Figure~\ref{Figure8}. The $\delta/a$ ratio 
is $<$1\% over inner  $10\arcsec$ ($\sim$4\,kpc),  after this radius the ratio
increases up to an upper limit of 15\% ($\delta=\sim$1.2\,kpc). Displacements of this order, $10\%-20\%$, 
has been  observed in interacting ellipticals and are not present in isolate systems \citep[e.g.][]{Lauer1986,Lauer1988,Davoust1988,Gonzales-Serrano2000}. The off-centering isophotes are the signature of m=1 tidal response mode, which can occur in an encounter when the pericenter is comparable or less than characteristic radius of the system \citep{Combes1995}.

The surface brightness profile is shown in the bottom panel of Figure~\ref{Figure8}. The de Vaucouleurs Law was fitted to the  data through Levenberg-Marquard method, the parameters obtained were $\mu_{\rm{Re}}=25.37\pm0.04$ mag/arcsec$^2$ and Re=6.81$\pm0.16$\,arcsec ($3.1$\,kpc). At the outermost radii, the surface brightness profile presents a faint excess of light with respect to the de Vaucouleurs law. This excess is clearly seen in the residuals
 starting from $12\arcsec$ ($\sim5.5$\,kpc) which achieved a maximum $\mu=$ 0.25 mag/arcsec$^2$ of deviation from the de Vaucouleurs law at the largest SMA. These bright envelopes are often seen in ellipticals with nearby companions (as it is the case for NGC\,2936), whereas in isolated galaxies these envelopes are absent \citep{Kormendy3_1977}. Moreover, this effect is observed when the dynamical time of the encounter is much shorter
than that of outermost parts of the galaxy, ``feeling'' a tidal impulse force rather than differential force \citep{Aguilar1985,Binney_dynamics}. 
The above tidal characteristics  found out in the elliptical galaxy of pair Arp\,142 shall be explained in detail in \ref{effects_on_2937}.

\subsection{GalMer model simulation of the system}
\label{sec:galmer}

\begin{figure}
\includegraphics[width=\columnwidth]{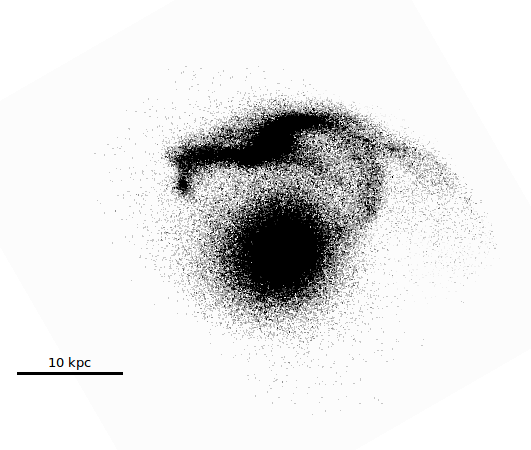}
\includegraphics[width=\columnwidth]{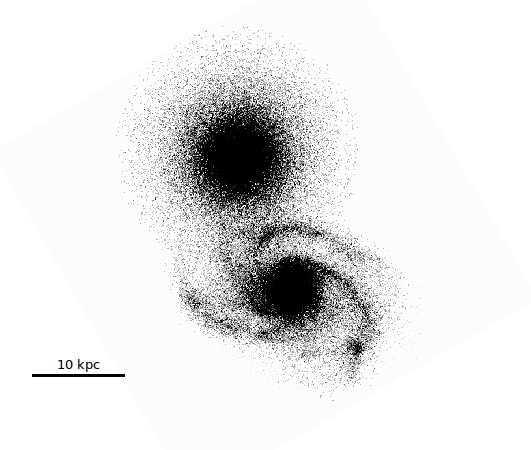}
\caption{ Snapshot of total mass (star+gas) for the simulated system. Top-panel: actual stage 50Myr after pericenter passage. Bottom-panel: pericenter passage.
The angles used for visualization were phi -28, theta 37 \citep[for more details see][]{Di-Matteo:2008qy}. } 
\label{Figure9}
\end{figure}

In order to build a dynamical scenario for the NGC\,2936/2937 system, we have explored   
the library of galaxy merger simulations provided by the  GalMer database \citep{2010A&A...518A..61C}.
In particular, we have { selected} a  major merger interaction (1:1 mass ratio) between a giant elliptical (E0) and spiral (Sa) {as the HST images of Arp 142 suggest.  Each simulation, with different  orbital parameters (initial orbital energy, angular momentum, direct
or prograde orbit, and {{ disc orientation}} with respect to orbital plane), was  visualized using the snapshot preview at different ages (in steps of 50 Myr, as it is provided by the database), changing its projection and, searching for the view that  best resemble the morphology of Arp 142.  Finally} the orbit that better matches the observed morphology of Arp\,142  is a prograde orbit with an initial velocity ($v_{ini}$) of 300\,km\,s$^{-1}$, a perigalacticum ($r_{peri}$) of 8\,kpc with  the disk of the spiral galaxy to be perpendicular to the orbit plane. From the simulation, we conclude that the current stage  of the system would be { $50\pm25$\,Myr} after the first pericenter passage. { The uncertainty comes  from the numerical simulation step of $50$\,Myr}. In Figure~\ref{Figure9} at  the top panel, we show a snapshot of the actual stage and at  the bottom a  snapshot of 50\,Myr before, at pericenter passage.
We can see that the first snapshot  reproduce  qualitatively very well  the main morphologic features of the studied system (c.f. Figure~\ref{Figure1}). The large plume at North-West direction and  the faint tidal bridge toward North direction (see Section \ref{bridge}). The perpendicular position of spiral disk  with respect to orbital plane explains why the tidal arms are bending toward the elliptical galaxy producing these features that give the famous ring-like form of the Penguin system. By observing the pericenter snapshot, we can conclude that these tidal features are ``new'', and  therefore  a maximum age of { $50\pm25$\,Myr}. 

\begin{figure}
\includegraphics[width=\columnwidth]{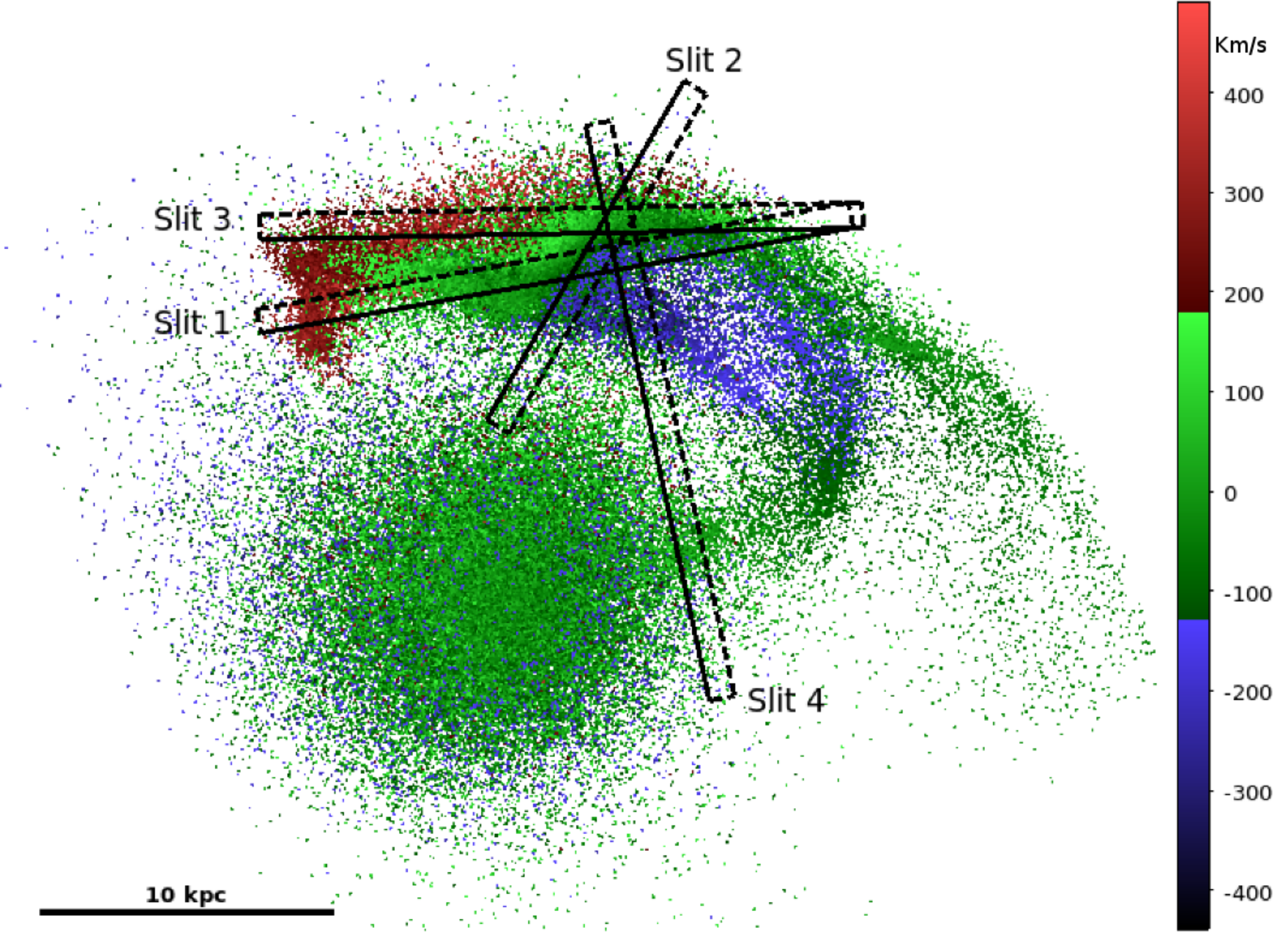}
\caption{Velocities map for the simulated system. See text for details. } 
\label{Figure10}
\end{figure}

\begin{figure}
\includegraphics[width=\columnwidth]{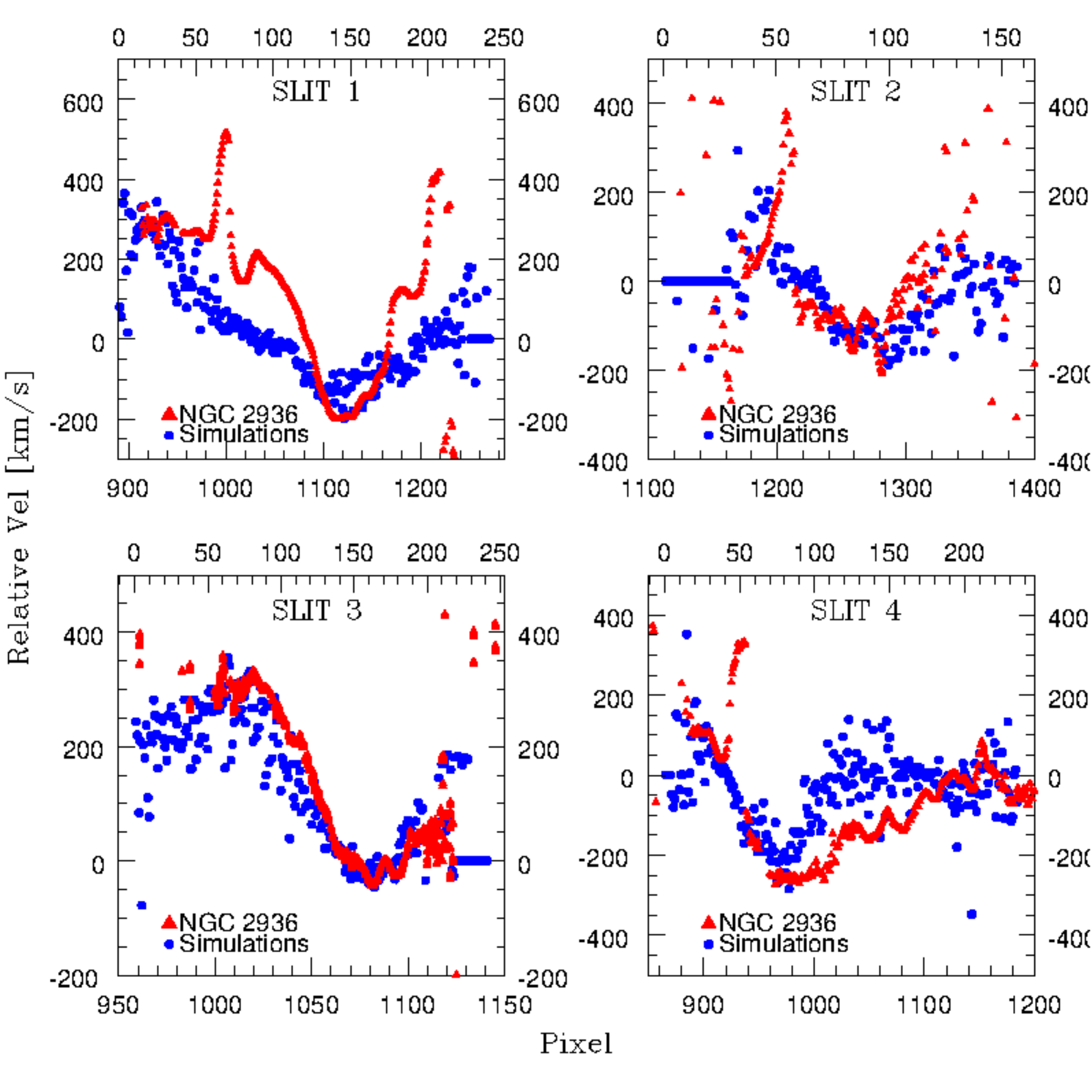}
\caption{Radial velocity profiles for the Arp\,142 system -  GalMer simulations and observations. The red (blue) profiles are the observed (simulated) radial velocities along the four slits. On each panel lower x-axis corresponds to the GMOS pixels while upper $x$ axis corresponds to the  simulated pixel silts from Fig. \ref{Figure10}. $y$ axis corresponds to the relative velocities for simulations and observations. NGC\,2963 velocity curves for the observations were derived by cross-correlating one spectrum (i.e. one row) of the extended 2D spectra against all the remaining  ones of  the same slit. }
\label{Figure11}
\end{figure}

One could also attempt to compare the kinematics of the system in pseudo slits similar to those used for the real observations. Figure~\ref{Figure10} shows the output of the GalMer simulation for the velocity field of the system.  We obtained the radial velocity profiles along slits in similar positions to those used in the real observations and these are plotted in Figure~\ref{Figure11}, together with the corresponding observed velocities. Both velocity profiles show similar shapes when the model and the observations are compared.
We stress that this is just an exercise to show that it is possible that the elliptical galaxy NGC\,2937 is, in fact, the perturbed of the NGC\,2936 spiral galaxy. However, given the nature of the problem we cannot assure that this is the only configuration that will describe the interacting scenario for this system. 

\begin{figure}
\includegraphics[width=\columnwidth]{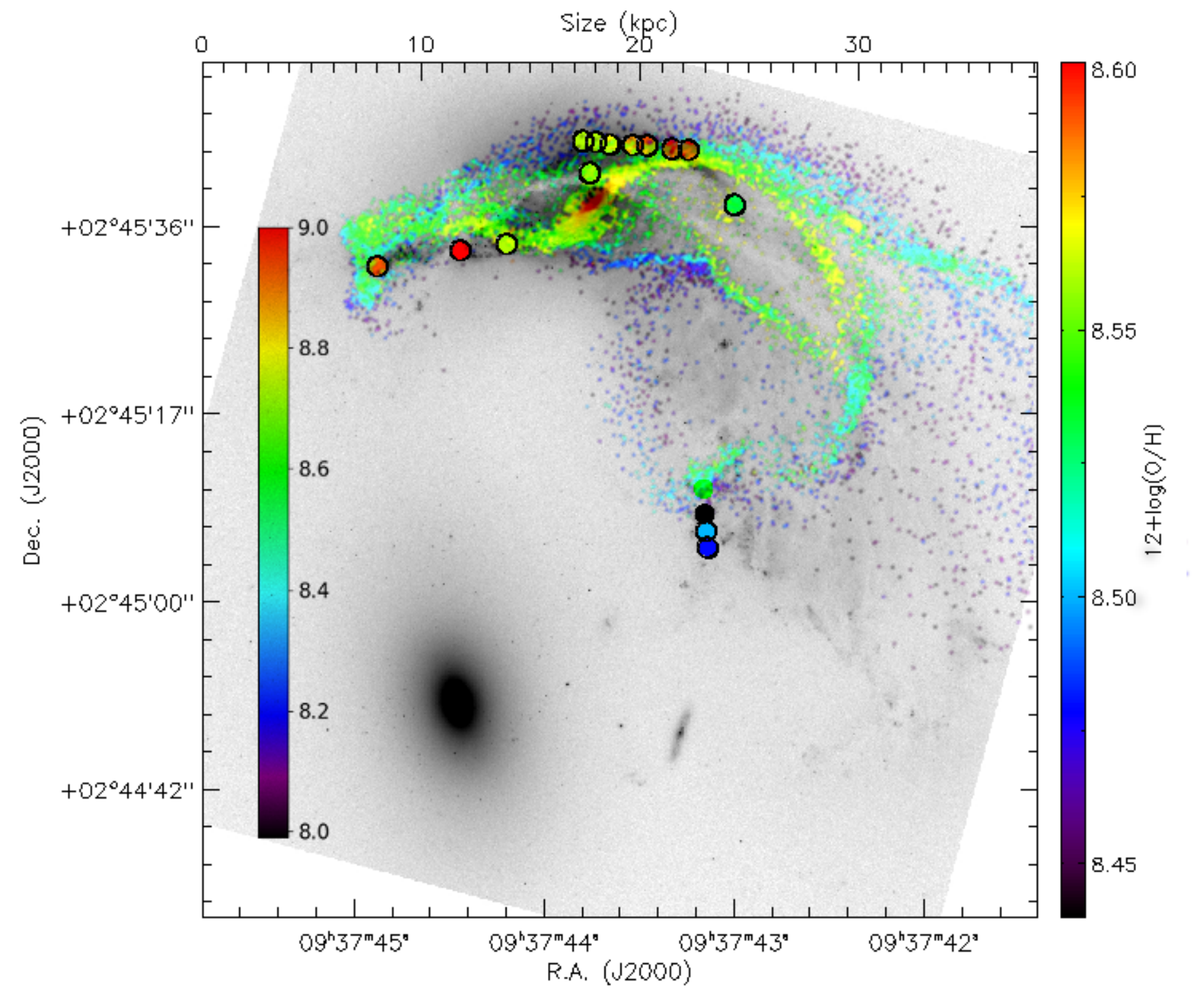}
\caption{{{GalMer color-coded spatial distribution of the abundances 
overlaid and scaled with  abundances showed on 
\ref{Figure4}. Big filled dots correspond to the measured regions while small dots are the output from GalMer simulations. For comparison between model and observations colors bar use the same cmpa, but different scale.  Right bar corresponds to dots, while left bar corresponds to Galmer simulation, same units as right bar.}}. 
}
\label{Figure_OH}
\end{figure}

{{Since GalMer {simulations} also provides metallicity {evolution of the system}, we can further extend this exercise to observe its expected distribution. Since the metallicity estimations from the observations are discrete (i.e. few selected regions), we have produced  in Fig. \ref{Figure_OH} a schematic comparison between the observations and the model. We stress that this is a rough  comparison since GalMer results were normalized in order to match the size of the background image.  Considering these caveats the trend predicted by the model show that  lowest metallicites are seen on the plunge while the  highest metallicities concentrate in the galaxy nucleus. Due to the uncertainties in our data, it is not possible to conclude if there is a metallicity gradient as the model shows, however the observed values seems to reproduce the relative values according to the position on the galaxy, i.e., high values close to the nucleus, low values in the plunge.}}

\section{Discussion}

\label{Discussion}

\subsection{Arp\,142: A peculiar S+E pair of galaxies}

Arp\,142 is an excellent example of the morphological transformation produced by a gravitational encounter. The southern-east side of the galaxy NGC\,2936 displays an extended structure, which was most probably produced by the tidal field of the elliptical galaxy NGC\,2937. In this sense, the spiral structure of NGC\,2936 seems to be clearly modified due to the interaction with its companion. In general, galaxy transformation in pairs of galaxies has been mainly studied in S+S pairs. However, there are few studies focused on the properties of S+E pairs \citep{Kojima:1997qb,Xu:2010ix,Jutte:2010zp,Hernandez-Ibarra:2016yt}, despite their importance in the mass assembly scenario. For instance, \cite{Xu:2010ix} studied a sample of 12 elliptical-spiral pairs. On average, these authors found no enhancement in the sSFR of these pairs. This is not the case for the intriguing system Arp\,142. Considering their values, (see Table~\ref{tableabundances_arp142}) the spiral member of Arp\,142 has a {$\log{sSFR}$ of $\sim$\,-10.51}, implying a SFR of 8.9 M$_{\odot}$ yr$^{-1}$ (SFR determined only from IR luminosities). This value is even larger if we take into account the total IR luminosity listed by these authors (as described in section \ref{colorimage}). This indicates that NGC\,2936 has an enhanced sSFR. Moreover, by using the H$\alpha$ luminosity, we found a total SFR of 35.9 M$_{\odot}$ yr$^{-1}$ \citep{Calzetti:2000fh}{{, or 17.5M$_{\odot}$ yr$^{-1}$ \citep{Fitzpatrick:1999ca}. Taking into account both values we find that they are}} larger than typical SFRs for spiral galaxies  (e.g. 1.17 M$_\odot$ yr$^{-1}$ from \citet{Xu:2010ix}, 2.6 M$_\odot$ yr$^{-1}$ from \citealt{Darg:2010mf}). {In this sense, what is the reason for this enhancement in SFR, considering that this enhancement is not the typical case for spirals in S+E pairs, as shown by \citet{Xu:2010ix}, and more recently by \citet{Cao:2016gn}. Indeed, in the case of S+E pairs, \citet{Hwang:2011rp} suggest that the X-ray emission associated with an elliptical galaxy could suppress the SFR in its companion spiral, but this seems to not be the case for NGC\,2936, probably due to the mass of the elliptical companion, which is not high enough to suppress the star formation in NGC\,2936. Actually, the interaction process between NGC\,2936 and NGC\,2937 could had produced gas flows and gas compression which may be the reasons of the star burst event in the eastern side of NGC\,2936. In this context, previous authors have found that close galaxy pairs display an enhancement in their SFRs (\citet{Ellison:2008rc}). Despite this is not the typical case for S+E pairs, the intriguing system Arp\,142 seems to be an exception. New 3D spectroscopy (in the optical and submm) can be very useful in order to disentangle the mechanisms that favour the star formation activity in this system.}


\subsection{The effects of the interactions on NGC\,2936}

The interaction that is taking place in Arp\,142 is strongly affecting and transforming the spiral NGC\,2936. Moreover, NGC\,2936 does not display the typical metallicity distribution expected for a spiral galaxy of its mass. Indeed, abundances show no gradient (within the error) and we interpreted it as a result of the re-distribution of the gas in the galaxy disk due to the inflow from the outskirt to the center of the galaxy. This gas mixing process  has been observed in interacting galaxies \citep[e.g.][]{Kewley:2006pi,Rupke:2010bc,Torres-Flores:2014bs,Rosa:2014al,Olave-Rojas:2015wj} %
Numerical simulation predicted the flattening of the metallicity gradient to  occurs just after first pericenter passage \citep{Rupke:2010bc,2011MNRAS.417..580P} hence the age of interaction stage, about  { $50\pm25$ Myr} after pericenter, that we estimated (Sec \ref{sec:galmer})  
is a good agreement with this scenario. The presence of the AGN and the evidence of the central region being ionized in NGC\,2936  hinted by the BPT diagram  agrees with  \citet{Ellison:2011it}, and thus with  the  higher fraction seen in close pairs of composite galaxies (i. e. galaxies whose central region is ionized by nuclear activity and massive stars) than in isolated systems.  A not surprising result since a large fraction of AGNs have been found  in close pairs compared with isolated galaxies \citep{Satyapal:2014cg}. However,  our results were obtained by using long-slit technique, which prevents us a spatially detailed analysis of the ionization in the center of the galaxy. Future IFU observations will elucidate the origin of the ionized nuclear activity in NGC\,2936.

\subsection{The effects of the interaction on NGC\,2937}
\label{effects_on_2937}

The analysis of NGC\,2937 (section \ref{sec:iso}) shows that isophotes appear to have a common center until  4 kpc ($\sim$9"). Then the  centers of outermost isophotes move 1.2 kpc toward the North in the same direction of the faint tidal bridge that connects both galaxies, whereas the inner part of NGC\,2937 is moving {\bf{towards}} the South. On the other hand, at large radii (from $\sim$13") the surface brightness profile shows an increasing enhancement outward
with respect to de Vaucouleurs law (Figure.~\ref{Figure8}), although small in magnitude. Both phenomena were explained 
by \citet{1986ApJ...307...97A} through N-Body simulation of spherical galaxies having a tidal encounter. 
The emergence of these kind of tidal perturbations, depending on whether the internal dynamical time, $t_{inner}=2r/\sigma$, is lesser or larger than the encounter dynamical time, $t_{encounter}=b/v$  (where $r$ is the radius of inner part, $\sigma$ is star velocity dispersion, $b$ the parameter of impact, and $v$ is the relative velocity of approximation). When  $t_{encounter}\geq t_{inner}$ the particles inside $r$ have enough time to react to the perturbation, and therefore the tidal approximation applies, the inner region is elongated toward the radial direction of center-of-masses, and (as a whole) is pushed away on the same direction,  presenting  an offset with respect to the outer envelope \citep[c.f. see panels (b) and (c) of Figure. 1 in][]{1986ApJ...307...97A}. Hence the inner part of NGC\,2937 moved in the North-South direction  toward closest passage (see bottom-panel in Figure.~\ref{Figure9}).
The inner region returns to its normal position because of dynamical friction. This displacement lasts around $1-3\times10^8$\,yr \citep{Combes1995}. On the other hand, at the outer regions the $t_{encounter}\leq t_{inner}$, therefore the impulsive approximation applies: the particles barely change its potential energy, thus the energy of the encounter is transferred in the form of kinematic energy to outer particles ``heating" them  \citep{1986ApJ...307...97A}, yielding into an expansion of the outer regions and producing an enhancement of the surface brightness profile at these radii, as it is seen in Figure. 2 of \citet{1986ApJ...307...97A}. The tidal distension lasts around $0.5\times10^9$\,yr after pericenter passage.

One could test if these tidal approximations are valid for the elliptical galaxy of the Arp\,142 system by calculating $t_{inner}$ and $t_{encounter}$. In order to do so, we consider the $\sigma=273$\,km\,s$^{-1}$ obtained from data release 12 of Sloan Digital Sky Survey \citep[SDSS;][]{2000AJ....120.1579Y}. $\sigma$ was calculated following  the method described in \citet{2012AJ....144..144B}. So, by taking $r$ as the radius until we observed the off-centric isophotes and where begin the enhancement of surface bright profile, around 4\,kpc, we obtain a $t_{inner}\simeq2\times4$\,kpc/273\,km\,s$^{-1}\simeq2.8\times10^7$\,yr. An approximation of $t_{encounter}$ can be obtained   from numerical simulation (Section \ref{sec:galmer}), then $t_{encounter}=8$\,kpc/300\,km\,s$^{-1}=2.6\times10^7$\,yr. The  values for $t_{inner}$ and $t_{encounter}$ are quite similar, which means that the both tidal approximations explained very well the tidal features observed in NGC\,2937.


\subsection{The future of Arp\,142}    

Because in this work we are presenting the properties of the distorted spiral  NGC\,2936,   triggered by the interaction with the elliptical NGC\,2937,  a natural step is to further explore the interacting system Arp\,142 since spiral-elliptical (gas/dust rich) interactions are one of the avenues on which elliptical galaxies may accrete dust, yielding to an elliptical (or early type) galaxy with dust lines, with a patchy or filamentary dust distribution \citep{Goudfrooij:1994kx}  as it is the historical  case of NGC\,5128 \citep{Baade:1954qy}. Moreover, \citet{Kaviraj:2012ys} showed  that $\sim$ 65\% of ellipticals (early-type galaxies) that contain dust present distorted morphologies, concluding that dust is likely to be the result of the recent mergers. They also found that these systems are located in low-density environment with  80\% of them  inhabiting in the field. In the context of these results, the isolated pair Arp\,142 seems to be a good candidate for a future dusty early-type galaxy. {GalMer simulations predict that the system will merge at an age of $\sim$ 3200\,Myr. Unfortunately GalMer simulations do not include dust as a separate component and therefore we cannot give numerical proofs  that the final stage of the pair will be a dusty elliptical.}  Whether the interacting pair NGC\,2936-NGC\,2937 will become an early type dusty galaxy remains to be further explored by numerical simulations based on future follow-ups.


\section{Conclusions}
\label{Conclusions}

We have analyzed GMOS spectroscopy of selected star-forming regions of NGC\,2936, member of the interacting pair Arp\,142. Our results  show an enhanced  star formation rate, probably triggered by the interaction with NGC\,2937. Star-forming regions in NGC\,2936 display oxygen abundances consistent with the solar value and electron densities larger than the typical values found in non-interacting systems. In addition, our extinction measurements are in agreement with the position of the H{\sc ii} regions on the optical images, like the regions near the dust line seen in NGC\,2936. 
Finally, we also find evidence suggesting that the nucleus of NGC\,2936 is being ionized by AGN activity. Regarding the elliptical member of the pair, isophotal analysis unveils the tidal effects on NGC\,2937 showing non-concentric isophotes at inner radii, besides that at large radii the surface brightness profile deviates from de Vaucouleurs law. Using  GalMer simulations  we are able to reproduce the observed radial velocity profiles  and observed galaxy morphologies only considering the interaction of an elliptical and a spiral galaxy, discarding, as a first approach a third member on the interaction.
 The current stage of the system would be about { $50\pm25$\,Myr} after the first pericenter passage. The North-West plume of  NGC\,2936 and the North faint tidal bridge between galaxy would be the result of 
that initial configuration of the system: the perpendicular orientation of the spiral disk with respect to the orbital plane.
The maximum age estimated for these tidal structures is { $50\pm25$\,Myr}.
We have proved that the third galaxy UCG\,05130\,NOTES01 have a difference in radial velocity large enough to do not be part of the interaction of the Arp\,142. 
To conclude, we believe that Arp\,142 is a  candidate to become a dusty spiral at the final stage of the merger.


\section*{Acknowledgements}
{{Authors thank the referee,  Javier Piqueras L\'opez, for his useful comments and suggestion that greatly improved this paper. }}
M.~D.~Mora acknowledges CONICYT, Programa de astronom\'ia, Fondo GEMINI-CONICYT: Este trabajo cont\'o con el apoyo de CONICYT, Programa de Astronom\'ia, Cargo de Astr\'onomo de Soporte GEMINI-CONICYT 2018. S. Torres-Flores acknowledges the financial support of Direcci\'on de Investigaci\'on y Desarrollo de la ULS, through a project DIDULS Regular, under contract PR16143. 
 V. Firpo acknowledges support from CONICYT Astronomy Program-2015 Research Fellow GEMINI-CONICYT (32RF0002). F. Urrutia-Viscarra acknowledges the financial support of the Chilean agency Conicyt $+$ PAI/Concurso nacional apoyo al retorno de investigadores/as desde el extranjero, convocatoria 2014, under de contract 82140065. CMdO acknowledge funding from FAPESP (program 2009/54202-8) and CNPq.  J.\,A.\,H.\,J. thanks to Brazilian  institution CNPq for financial support through  postdoctoral fellowship (project 150237/2017-0) {and Chilean institution CONICYT, Programa de Astronom\'ia, Fondo ALMA-CONICYT 2017, C\'odigo de proyecto 31170038}. Based on observations obtained at the Gemini Observatory, which is operated by the Association of Universities for Research in Astronomy, Inc., under a cooperative agreement with the NSF on behalf of the Gemini partnership: the National Science Foundation (United States), the National Research Council (Canada), CONICYT (Chile), Ministerio de Ciencia, Tecnolog\'{i}a e Innovaci\'{o}n Productiva (Argentina), and Minist\'{e}rio da Ci\^{e}ncia, Tecnologia e Inova\c{c}\~{a}o (Brazil).
Based on observations made with the NASA/ESA Hubble Space Telescope, obtained from the Data Archive at the Space Telescope Science Institute, which is operated by the Association of Universities for Research in Astronomy, Inc., under NASA contract NAS 5-26555. These observations are associated with program \# 12812.

\bibliographystyle{mnras}
\bibliography{ARP_142_Accepted_14_Jun}



\end{document}